\documentclass[11pt]{article}

\usepackage[utf8]{inputenc}
\usepackage{mathtools}
\usepackage{amssymb, amsfonts, amsthm}
\usepackage{enumitem}
\usepackage{physics}
\usepackage[toc,page]{appendix}
\usepackage{authblk}
\usepackage{boldline, multirow}
\usepackage[margin=1in]{geometry}
\usepackage{parskip}
\usepackage{microtype}
\usepackage{colortbl}
\usepackage[font=small,labelfont=bf]{caption}
\usepackage[section]{placeins}
\usepackage{graphicx}
\usepackage{xcolor}
\usepackage{url}
\usepackage{braket}
\usepackage{float}
\usepackage{booktabs}
\usepackage{tabularx}
\usepackage{array}
\usepackage{xspace}

\usepackage[
    backend=biber,
    style=numeric,
    sorting=none
]{biblatex}
\usepackage[colorlinks=true, linkcolor=blue, citecolor=blue, urlcolor=blue, unicode=true]{hyperref}
\usepackage{cleveref}

\theoremstyle{definition}

\theoremstyle{remark}

\newcommand{\model}{FlowMeas\xspace}
\newcommand{\contactemail}[1]{%
  \thanks{Contact author: \href{mailto:#1}{\texttt{#1}}}}

\title{Generative Learning for Quantum Measurement Design}
\author[1,2]{Jun Dai\contactemail{jun.dai@mila.quebec}}
\author[3]{Olivier Nahman-L\'{e}vesque}
\author[1,2,6]{Guillaume Rabusseau}
\author[4]{Hong-Ye Hu\contactemail{hongyehu.physics@gmail.com}}
\author[1,3,5]{Cunlu Zhou\contactemail{Cunlu.Zhou@USherbrooke.ca}}

\affil[1]{Mila – Québec AI Institute, Montréal, QC, Canada}
\affil[2]{Département d’informatique et de recherche opérationnelle, Université de Montréal, Montréal, QC, Canada}
\affil[3]{Institut quantique, Université de Sherbrooke, Sherbrooke, QC, Canada}
\affil[4]{Department of Physics, Harvard University, Cambridge, MA, USA}
\affil[5]{Department of Computer Science, Université de Sherbrooke, Sherbrooke, QC, Canada}
\affil[6]{CIFAR AI Chair}

\date{}

\begin{document}

\maketitle

\begin{abstract}
Extracting quantum information from a quantum state is a fundamental task of quantum computation, often requiring the estimation of many non-commuting observables under a finite measurement budget. For both near-term and early fault-tolerant settings, the measurement protocol must balance statistical efficiency against implementation resources such as circuit depth, connectivity, and entangling-gate count. Many existing strategies focus on two extremes: hardware-friendly product measurements with high sampling cost, and fully commuting measurements with deep circuits. Here we recast resource-constrained measurement design as a generative learning problem. We introduce \model, which uses a generative flow network to directly sample finite ensembles of shallow Clifford measurement circuits subject to a prescribed shot budget and hardware constraints. At zero entangling depth, \model learns qubit-wise commuting measurement schedules and already matches or improves leading product-measurement methods on nearly all molecular benchmarks. Allowing one or two entangling gate layers yields further reductions in energy estimation error of up to $27\%$ relative to the strongest state-independent product-measurement baseline. The learned policy can also be reused across related Hamiltonians, substantially accelerating retraining along a molecular potential-energy surface. We further obtain results for molecular Hamiltonians with up to 20 qubits and apply the framework to a compactly encoded 54-qubit interacting fermionic model, extending the demonstrated scale beyond prior molecular benchmarks. These results establish generative learning as a flexible and unified framework for quantum measurement design under practical resource constraints.
\end{abstract}

\section{Introduction}
Extracting useful quantum information from a quantum state is a fundamental task in quantum information processing. The target may be an energy required by a near-term quantum algorithm~\cite{peruzzo2014variational,preskill2018nisq,tilly2022variational,patel2025quantum}, a collection of correlation functions, a learned Hamiltonian~\cite{huang2023learning}, or a certificate of state preparation or device performance~\cite{huang2020predicting,carrasco2021verification}. In all these settings, the quantum state is accessed through a finite number of measurements, and the choice of measurement circuits can dominate the overall cost~\cite{wecker2015progress,gonthier2022measurements}. Measurement design is therefore simultaneously a statistical and an implementation problem: one must reduce the number of state preparations and shots while respecting constraints on circuit depth, gate set, connectivity, and fidelity.

Many observables of interest decompose into sums of Pauli operators. A measurement circuit can provide information about every term that it maps to a product of Pauli-$Z$ and identity operators before computational-basis measurement. A single outcome can therefore be reused to estimate several compatible terms. This reuse can substantially reduce the sampling cost~\cite{bonet2020nearly}, but jointly diagonalizing commuting Pauli terms that are not qubit-wise commuting (QWC) often requires additional entangling gates and deeper circuits. Measurement design consequently exhibits a basic depth--sampling trade-off.

Most established methods focus on one of the two limiting regimes. QWC strategies use only independent single-qubit basis rotations and are therefore inexpensive to implement, but their restricted measurement family may require many shots~\cite{hadfield2022measurements,huang2021efficient,wu2023overlapped, verteletskyi2020measurement}. Fully commuting (FC) strategies can combine much larger sets of observables~\cite{izmaylov2020unitary,yen2020measuring,zhao2020measurement,huggins2021efficient,yen2021cartan,crawford2021efficient}, but diagonalizing a general commuting family can require a Clifford circuit whose depth grows with system size and whose two-qubit gate count can scale as $O(n^2/\log n)$~\cite{gottesman1998heisenberg,aaronson2004improved,bravyi2021hadamard}. These limiting regimes are attractive partly because their compatibility conditions induce graph-coloring or clique-cover formulations ~\cite{verteletskyi2020measurement,yen2020measuring}, for which efficient heuristics are sometimes available despite the underlying NP-hardness. The practically relevant regime lies between these extremes: one would like to exploit a limited amount of entanglement while enforcing a maximum depth, a device connectivity graph, a native gate set, or a fixed number of measurement circuits. Such restrictions cannot be similarly mapped onto a graph problem, and thus measurement design in this intermediate regime is currently underexplored.

Randomized measurement protocols such as classical shadows offer a complementary route, and adapting the measurement ensemble to a given observable set is known to improve their sampling efficiency~\cite{elben2023randomized,garcia2021learning,zhao2021fermionic}. Random shallow Clifford measurements further provide bounded-depth protocols, but an unstructured random ensemble is not tailored to a specified observable set~\cite{bertoni2024shallow,ippoliti2023operator,hu2023classical,akhtar2023scalable,hu2025demonstration}. A closely related
prior method in this intermediate regime is the recently introduced derandomized shallow shadows (DSS) method~\cite{kirk_derandomized_2024}. DSS begins with probabilistic shallow-circuit ansatzes and fixes their gates sequentially. At each step, it evaluates the probability that a partially randomized candidate diagonalizes each target Pauli string. Tensor-network techniques make these probabilities computable with polynomial resources at bounded or polylogarithmic depth, but the resulting optimization is operationally indirect: conditional costs are repeatedly evaluated inside nested loops over measurement circuits, gates, and candidate assignments. The search is also greedy, committing sequentially to local gate choices that are not guaranteed to produce a globally optimized measurement ensemble.

\begin{figure}[htbp]
\centering
\includegraphics[width=0.95\textwidth]{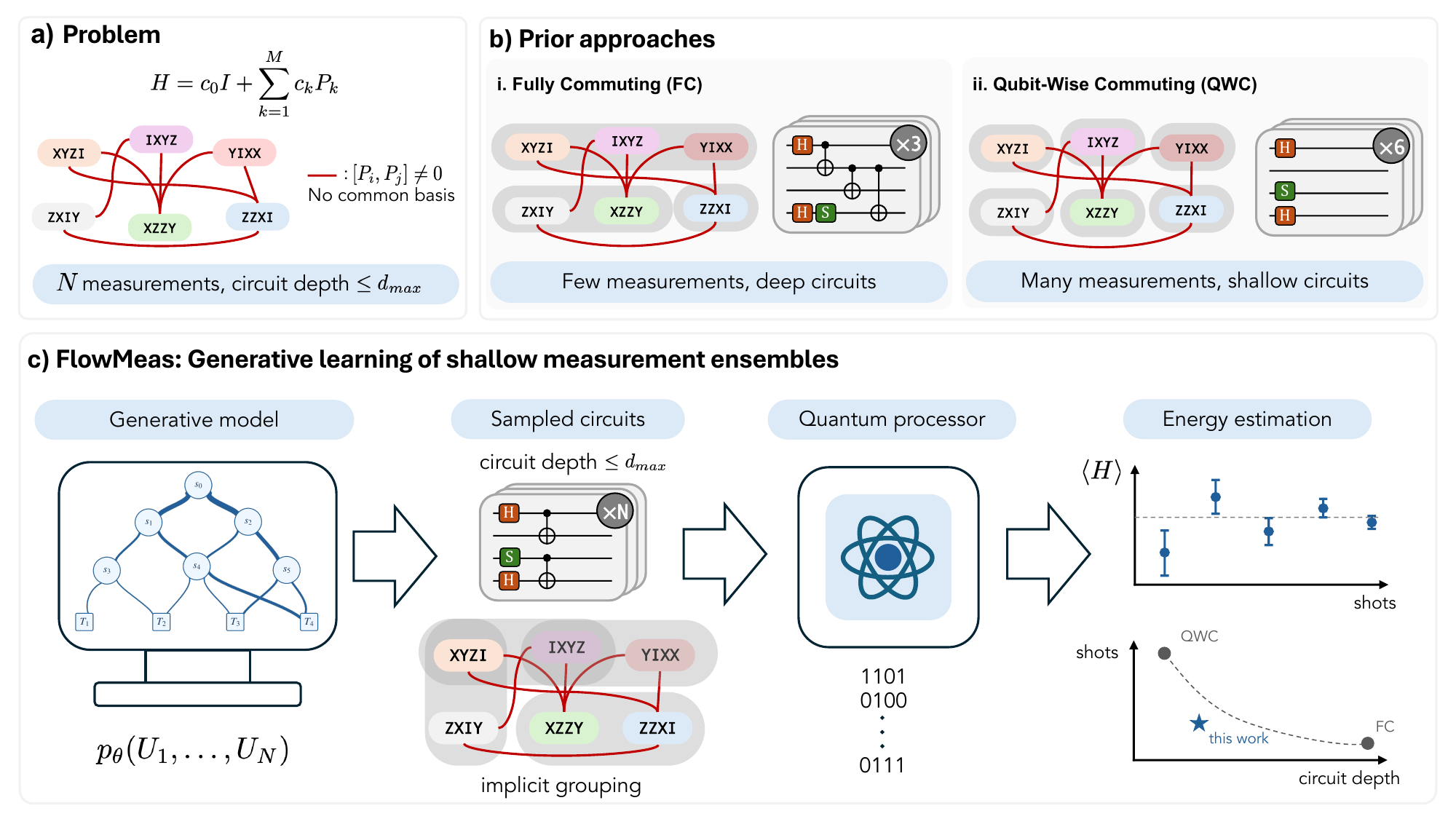}
\caption{\textbf{Resource-constrained quantum measurement design.} (a) Estimating $\langle H\rangle$ for $H=c_0I+\sum_k c_kP_k$ requires measuring many non-commuting Pauli terms under a finite shot budget and a hardware-limited circuit depth. (b) QWC measurements use only single-qubit basis rotations but  may require many distinct tensor-product Pauli measurement bases, each implemented as a separate circuit, whereas 
FC grouping can reduce the sampling cost at the price of deeper entangling circuits. (c) \model learns a generative model
$p_\theta(U_1,\ldots,U_N)$ over ensembles of $N$ shallow Clifford measurement circuits, each satisfying the prescribed depth and hardware constraints.  A sampled ensemble defines an executable protocol and an implicit overlapping grouping: one circuit can cover several Pauli terms, and one term can be covered repeatedly by several circuits. By varying the allowed circuit family, \model directly explores the depth--sampling trade-off and favors ensembles that combine shallow implementation with high statistical efficiency.}
\label{fig:overview}
\end{figure}

Here we introduce \model, a generative learning framework for resource-constrained measurement design. Rather than starting from a random ensemble and gradually derandomizing it, we learn a policy whose outputs are already executable measurement circuits. Given the target observables and their weights, a measurement budget, and an allowed circuit family, the model directly samples finite ensembles of deterministic shallow Clifford circuits. Their quality is evaluated collectively through exact Pauli coverage statistics. The design problem is thus transformed from explicit grouping or sequential derandomization into learning a distribution over hardware-compatible measurement circuits, as illustrated in Fig.~\ref{fig:overview}.

We instantiate \model using a generative flow network (GFlowNet)~\cite{bengio2021flow,malkin2022trajectory,bengio2023gflownet}.
GFlowNets construct discrete objects through sequences of actions and are trained, in the exact-flow limit, to sample completed objects with probability proportional to a reward. This is naturally suited to measurement design: circuits are compositional objects assembled from discrete gates, many structurally distinct circuits can have comparable utility, and an effective measurement protocol requires an ensemble with complementary rather than redundant coverage. Compared with standard reward maximizing reinforcement learning formulations~\cite{sutton2018reinforcement}, GFlowNets explicitly target a reward-weighted distribution over terminal objects rather than only maximizing the expected reward of the learned policy. This distributional objective encourages exploration of multiple high-quality modes while amortizing part of the search into a constructive policy that can generate new candidates and be reused across related instances. GFlowNets have been applied to a range of discrete design problems, including biological sequence design~\cite{jain2022biological}, multi-objective molecular design~\cite{jain2023multiobjective}, graph combinatorial optimization~\cite{zhang2023letflows}, robust scheduling~\cite{zhang2023robust}, Bayesian network structure learning~\cite{deleu2022bayesian}, and discrete probabilistic modeling~\cite{zhang2022discrete}. In the quantum setting, related work has applied GFlowNets to commuting partitions of Hamiltonian terms~\cite{huidobro2024gflownets} and the synthesis of variational quantum circuits~\cite{dai2025flowqnet}; here the terminal object is instead a hardware-constrained ensemble of measurement circuits, and the reward evaluates the collective measurement protocol.

Compared with previous approaches, our learning-based formulation offers several advantages. Hardware and resource constraints are incorporated directly through the allowed circuit family, including the maximum entangling depth, gate set, connectivity graph, measurement budget, and costs assigned to different measurement settings. The method directly optimizes executable measurement ensembles, including their measurement multiplicities, rather than first constructing explicit groups and subsequently optimizing shot allocations. Crucially, every sampled candidate is a deterministic Clifford circuit, so its coverage of the target Pauli operators can be computed exactly by stabilizer tableau propagation~\cite{gottesman1998heisenberg,aaronson2004improved}. We implement these operations using GPU kernels, enabling efficient evaluation of ensemble hit counts and state-independent proxy costs. In contrast, DSS evaluates the diagonalization probabilities of partially randomized circuits using tensor-network methods \cite{kirk_derandomized_2024} during sequential gate fixing, and this calculation must be repeated across circuits, gates, candidate assignments, and target Pauli operators. Finally, the training objective depends only on the target observables, their coefficients, the generated circuits, and their coverage statistics, and therefore requires no prior knowledge of the measured quantum state. This is particularly useful when the state is not known before measurement, as in ground-state estimation and many other settings.

The resulting framework interpolates across standard measurement families. At zero entangling depth, it learns a finite QWC measurement schedule, including both the bases and their multiplicities. Increasing the allowed depth enlarges the family to shallow entangling Clifford measurements and, at sufficient depth, toward circuits capable of diagonalizing general FC groups. Numerically, the learned QWC schedules already match or improve leading product-measurement strategies on nearly all molecular benchmarks. One or two entangling layers provide additional gains, with reductions in energy estimation error of up to $27\%$ relative to the strongest state-independent product-measurement baseline. Direct comparisons also show that FlowMeas matches or improves DSS on four of the five shared molecular systems. Beyond accuracy, our direct generative approach extends the demonstrated problem size: we obtain results for molecular Hamiltonians with up to 20 qubits, whereas prior state-of-the-art molecular benchmarks were largely limited to at most 16 qubits. We further show that a policy trained at one molecular geometry can be reused across a potential-energy surface, reducing the required optimization iterations by factors ranging from approximately three to more than ten while preserving the final accuracy. Finally, we apply the method to compactly encoded spinless Hubbard models with up to 54 qubits, demonstrating its computational reach well beyond the systems accessible to exact state-vector simulation.

Overall, our work provides a flexible and unified framework for practical measurement design on resource-constrained quantum devices. As quantum processors improve, the measurement layer will remain subject to finite shots, imperfect gates, limited connectivity, and depth constraints. The relevant question is not only how to minimize measurement cost in isolation, but how to choose measurement circuits that jointly balance statistical efficiency and implementation resources. By using generative learning to optimize ensembles of shallow measurement circuits, our approach addresses this trade-off directly and opens a new route toward adaptive, hardware-aware measurement protocols for both near-term and early fault-tolerant quantum computing.

\section{Generative Learning of Measurement Circuits}
\label{sec:framework}

\begin{figure}[htbp]
\centering
\includegraphics[width=0.95\textwidth]{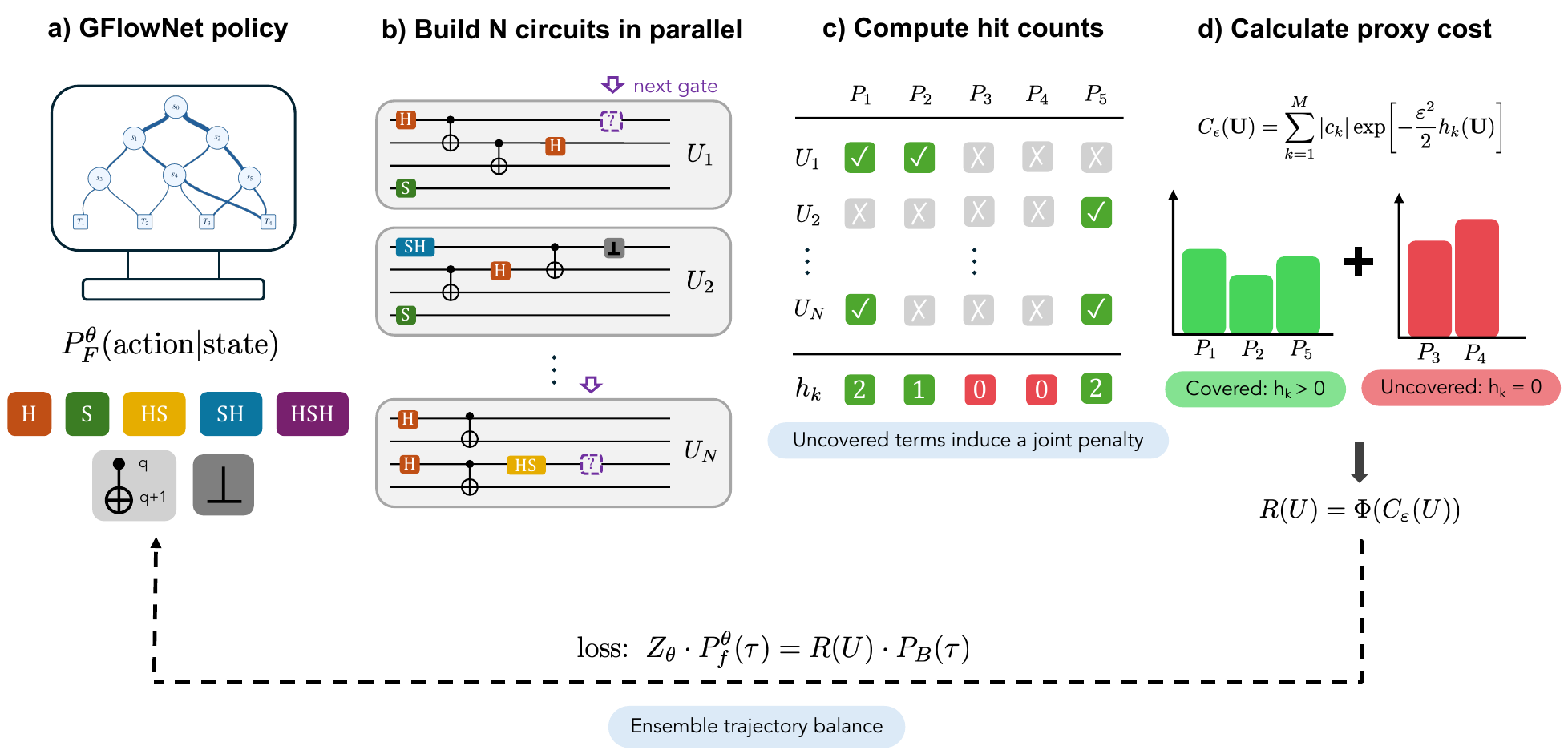}
\caption{\textbf{Generative learning of shallow measurement ensembles.}
(a) A shared GFlowNet forward policy $P_F^\theta(a\mid s)$ constructs Clifford measurement circuits from a discrete action set consisting of local Clifford gates, nearest-neighbor CNOTs, and a stop action $\bot$.
(b) The policy constructs $N$ circuit slots in parallel, with action masks enforcing the prescribed connectivity and CNOT-depth limit $d_{\max}$. The completed circuits form an executable ensemble
$\mathbf{U}=(U_1,\ldots,U_N)$.
(c) Each ensemble defines a binary hit matrix: an entry indicates that circuit $U_j$ maps Pauli term $P_k$ to a product of $Z$ and identity operators. The column sums give the hit counts $h_{\mathbf U}$, which quantify how the measurement budget is distributed across the Hamiltonian terms.
(d) The hit counts and Hamiltonian coefficients define a state-independent proxy cost $C(\mathbf U)$. The illustrated proxy decreases as important terms receive additional hits and penalizes terms left uncovered. A decreasing transformation produces the terminal reward $R(\mathbf U)=\Phi(C(\mathbf U))$, and the policy is trained using an ensemble-level trajectory-balance objective defined in Sec~\ref{sec:gflownet-objective}.
}
\label{fig:gfns}
\end{figure}

\model directly learns executable measurement circuits within a prescribed resource budget, as illustrated conceptually in Fig.~\ref{fig:overview} and algorithmically in Fig.~\ref{fig:gfns}. Consider a Hamiltonian
\begin{equation}
\label{eq:ham}
    H=c_0 I+\sum_{k=1}^{M} c_k P_k,
\end{equation}
where $\{P_k\}_{k=1}^{M}$ are the distinct non-identity Pauli strings and $c_k\in\mathbb{R}$ are their coefficients. Given a measurement budget of $N$ shots and an allowed family of shallow Clifford circuits, \model generates an ensemble
\begin{equation*}
    \mathbf{U}=(U_1,\ldots,U_N).
\end{equation*}

Each circuit corresponds to one measurement shot; repeated circuits therefore represent repeated measurements in the same basis. The ensemble jointly specifies both the measurement bases and their multiplicities, without requiring a separate grouping or shot-allocation procedure.

A circuit $U_j$ covers a Pauli operator $P_k$ if
$U_jP_kU_j^\dagger\in\pm\{I,Z\}^{\otimes n}$. We encode this condition
through the coverage indicator $\chi_{jk}$ and define the corresponding
ensemble hit count by
\begin{align}
    h_k(\mathbf U)
    &=
    \sum_{j=1}^{N}\chi_{jk},
    \label{eq:hit-count}
    \\
    \chi_{jk}
    &=
    \mathbf 1\!\left[
        U_jP_kU_j^\dagger\in\pm\{I,Z\}^{\otimes n}
    \right].
    \label{eq:hit-indicator}
\end{align}
A single circuit may cover many Pauli terms, while the same term may be covered repeatedly by several circuits. The resulting hit matrix therefore defines an implicit overlapping grouping of the Hamiltonian terms. Its column sums $h_k(\mathbf U)$ quantify how the finite measurement budget is distributed across the observable set.

FlowMeas uses a GFlowNet to generate these circuit ensembles, as illustrated in Fig.~\ref{fig:gfns}. Starting from empty circuits, a shared forward policy sequentially selects gates from a discrete Clifford action set. The $N$ circuit slots are constructed in parallel, subject to masks enforcing the allowed gate set, connectivity, and maximum entangling depth. The terminal reward is assigned only after the complete ensemble has been generated. Consequently, circuit utility is evaluated collectively: a circuit that covers observables missed by the rest of the ensemble can be more valuable than one that redundantly measures terms already receiving many hits. The GFlowNet objective trains the constructive policy to place greater probability on high-reward ensembles while retaining multiple distinct measurement protocols with comparable quality.

The terminal reward is obtained from a state-independent proxy cost $C(\mathbf{U})$,
\begin{equation}
    R(\mathbf{U})=\Phi\!\left(C(\mathbf{U})\right),
    \label{eq:proxy-reward}
\end{equation}
where $\Phi$ is a positive monotonically decreasing function. The proxy depends only on the Hamiltonian coefficients and the coverage statistics of the generated ensemble. It can therefore reward repeated coverage of important terms, account for the diminishing benefit of additional hits, and penalize terms that remain uncovered, without requiring information about the measured quantum state. Different proxy costs encode different statistical objectives; their definitions and empirical effects are studied in Secs.~\ref{sec:cost-functions} and~\ref{sec:proxy-results}.

Because every generated candidate is already a deterministic Clifford circuit, its complete Pauli coverage can be computed exactly by stabilizer tableau propagation. We implement these operations using GPU kernels, enabling efficient
evaluation of ensemble hit counts and state-independent proxy costs. Unlike DSS, which optimize partially randomized circuits, no probability distribution over intermediate circuits must be propagated, and no tensor-network methods are required to evaluate the terminal cost. This direct representation makes it possible to evaluate large batches of candidate ensembles efficiently during training.

Hardware constraints enter through the circuit generator. At zero entangling depth, the admissible circuits consist only of local Clifford rotations, and FlowMeas learns a finite QWC measurement schedule. Allowing one or more layers of nearest-neighbor CNOT gates enlarges the measurement family to shallow entangling Clifford bases. More generally, the action set and masks can encode a device connectivity graph, native Clifford gates, limits on entangling-gate count or depth, and nonuniform implementation costs. The framework therefore searches directly within the measurement family that can be executed on the target hardware.

This formulation differs from both explicit grouping and sequential derandomization. Grouping methods first construct compatible sets of observables and typically optimize their measurement allocations in a separate step. DSS instead starts from randomized shallow-circuit ansatzes and fixes their gates sequentially by repeatedly evaluating conditional costs. FlowMeas treats the measurement circuits themselves as the generated objects and explores their space through a learned distribution over constructive trajectories. The resulting protocol is direct, state-independent, and resource-aware, while retaining the flexibility to interpolate from product measurements to increasingly expressive entangling measurement families.

\section{Results}
\label{sec:numerical-results}

\subsection{Learning shallow measurement ensembles improves energy estimation}

We first investigate two distinct questions: whether generative learning improves the measurement schedule within a fixed product-measurement family, and whether allowing shallow entangling circuits provides additional benefit. We evaluate \model on ground-state energy estimation for eight Jordan--Wigner~\cite{jordan1928paulische} molecular Hamiltonians from the public \texttt{variances} collection~\cite{hadfieldVariances}, ranging from four to twenty qubits. The first six systems coincide with the shared molecular benchmark set used in previous comparisons, while the 20-qubit $\mathrm{C}_2$ and $\mathrm{HCl}$ Hamiltonians extend the demonstrated molecular scale beyond that set. Reference ground states are obtained by exact diagonalization. For each fixed measurement ensemble, we perform 500 independent shot-noise trials and report the root-mean-squared error (RMSE) of the estimated energy in Hartree.

We compare with published results for largest-degree-first grouping (LDF)~\cite{verteletskyi2020measurement}, locally biased classical shadows (LBCS)~\cite{hadfield2022measurements}, derandomized classical shadows (Derand)~\cite{huang2021efficient}, and overlapped grouping measurements (OGM)~\cite{wu2023overlapped}. These methods are state-agnostic, use product Pauli measurement bases and are evaluated under the same budget of 1000 measurements. For \model, the depth parameter $d_{\max}$ counts only CNOT layers; single-qubit Clifford rotations are not counted toward
$d_{\max}$. Thus, $d_{\max}=0$ restricts the search to QWC product measurements, whereas $d_{\max}=1,2$ admit shallow entangling Clifford bases.

\begin{table}[htbp]
\centering
\small
\setlength{\tabcolsep}{4.2pt}
\caption{\textbf{Ground-state energy RMSE for molecular Hamiltonians.} All values are in Hartree and use a budget of 1000 measurements. LDF, LBCS, Derand, and OGM values are published product-measurement results from Ref.~\cite{wu2023overlapped}. The shaded column highlights the product-measurement limit of \model. The remaining columns allow one or two CNOT layers. Boldface marks the lowest reported value in each row. ``n.r.'' indicates that the corresponding comparison was not reported.}
\label{tab:molecular-rmse}
\begin{tabular}{lccccccc}
\toprule
System & LDF & LBCS & Derand & OGM & \cellcolor{gray!12}$d_{\max}=0$ & $d_{\max}=1$ & $d_{\max}=2$\\
\midrule
$\mathrm{H}_2(4)$ & 0.019 & 0.043 & 0.018 & \textbf{0.011} & \cellcolor{gray!12}0.013 & 0.012 & 0.012\\
$\mathrm{H}_2(8)$ & 0.149 & 0.128 & 0.067 & 0.051 & \cellcolor{gray!12}0.042 & 0.038 & \textbf{0.037}\\
$\mathrm{LiH}(12)$ & 0.231 & 0.122 & 0.063 & 0.036 & \cellcolor{gray!12}0.036 & 0.036 & \textbf{0.035}\\
$\mathrm{BeH}_2(14)$ & 0.426 & 0.275 & 0.103 & 0.072 & \cellcolor{gray!12}0.063 & 0.061 & \textbf{0.060}\\
$\mathrm{H}_2\mathrm{O}(14)$ & 1.090 & 0.549 & 0.257 & 0.129 & \cellcolor{gray!12}0.109 & 0.108 & \textbf{0.104}\\
$\mathrm{NH}_3(16)$ & 1.063 & 0.484 & 0.225 & 0.151 & \cellcolor{gray!12}0.138 & \textbf{0.137} & 0.139\\
\midrule
$\mathrm{C}_2(20)$ & \multicolumn{4}{c}{n.r.} & \cellcolor{gray!12}0.322 & 0.216 & \textbf{0.211}\\
$\mathrm{HCl}(20)$ & \multicolumn{4}{c}{n.r.} & \cellcolor{gray!12}0.257 & 0.234 & \textbf{0.226}\\
\bottomrule
\end{tabular}
\end{table}

Table~\ref{tab:molecular-rmse} first isolates the effect of generative schedule design. At $d_{\max}=0$, \model improves on OGM for four of the six shared systems, matches it for $\mathrm{LiH}(12)$, and is slightly less accurate only for the four-qubit $\mathrm{H}_2$ instance. The gain therefore does not arise solely from access to entangling measurements: directly learning the product bases and their multiplicities already produces highly effective finite measurement schedules.

Allowing one or two CNOT layers provides a second source of improvement. The best \model result is lower than the strongest state-independent product-measurement baseline for five of the six shared systems. Relative to OGM, the RMSE is reduced by approximately $27\%$ for $\mathrm{H}_2(8)$, $17\%$ for $\mathrm{BeH}_2(14)$, $19\%$ for $\mathrm{H}_2\mathrm{O}(14)$, and $9\%$ for $\mathrm{NH}_3(16)$. The improvement is not strictly monotonic with depth in every finite training run, but every system benefits from at least one entangling configuration.

For the two 20-qubit Hamiltonians, corresponding published product-measurement comparisons are unavailable. Nevertheless, increasing the allowed depth from zero to two CNOT layers reduces the RMSE by approximately $34\%$ for $\mathrm{C}_2$ and $12\%$ for $\mathrm{HCl}$. These systems extend the molecular experiments beyond the 16-qubit limit of the shared published benchmark set. Earlier OGM and DSS molecular comparisons were conducted on systems of at most 16 qubits. 

\begin{figure}[htbp]
\centering
\includegraphics[width=0.86\linewidth]{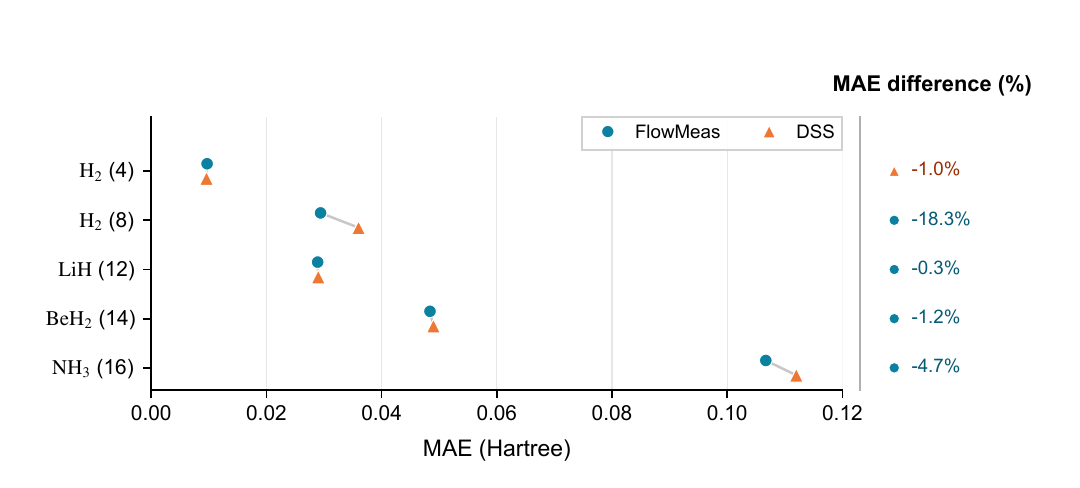}
\caption{\textbf{Direct comparison with derandomized shallow shadows.}
Mean absolute error (MAE) for five molecular Hamiltonians under a common budget of 1000 measurements and one CNOT layer. Both DSS and \model use the DSS confidence-based proxy objective. Orange filled triangles show DSS values from Ref.~\cite{kirk_derandomized_2024}, while blue filled circles show \model. The right column reports the relative MAE difference, normalized by the DSS MAE. \model matches or improves DSS on four of the five systems, with the largest reduction occurring for the eight-qubit $\mathrm{H}_2$ Hamiltonian.
}
\label{fig:dss}
\end{figure}

Because DSS reports mean absolute error (MAE) rather than RMSE, we present the direct comparison separately in Fig.~\ref{fig:dss}. The DSS values are from Ref.~\cite{kirk_derandomized_2024}. Under the same MAE metric, measurement budget, CNOT depth, and confidence-based proxy objective, \model matches or improves DSS on four of the five shared systems. The largest reduction is $18.3\%$ for $\mathrm{H}_2(8)$, followed by $4.7\%$ for $\mathrm{NH}_3(16)$. The LiH and $\mathrm{BeH}_2$ results are nearly equal, while DSS gives a $1.0\%$ lower MAE for the smallest $\mathrm{H}_2$ instance. 

\subsection{Effect of the state-independent training proxy}
\label{sec:proxy-results}

The final measurement ensembles are evaluated using ground-state energy RMSE, but this state-dependent quantity is in general unavailable during training. \model instead uses a state-independent proxy cost computed only from the Hamiltonian coefficients and the Pauli hit counts of a candidate ensemble. Different proxies encode different approximations to the statistical measurement cost and may favor different allocations of the finite measurement budget. Because they do not use the target state or its Pauli covariances, their ability to predict the final energy error must be assessed empirically.

We compare three training objectives at fixed CNOT depth
$d_{\max}=2$: a variance-plus-bias (VB) proxy, the confidence proxy used by DSS, and the OGM diagonal-variance proxy. All runs use the same circuit family, model architecture, measurement budget, and training protocol; only the terminal reward is changed. The resulting measurement ensembles are generated by \model in all cases.

\begin{figure}[htbp]
    \centering
    \includegraphics[width=0.8\linewidth]{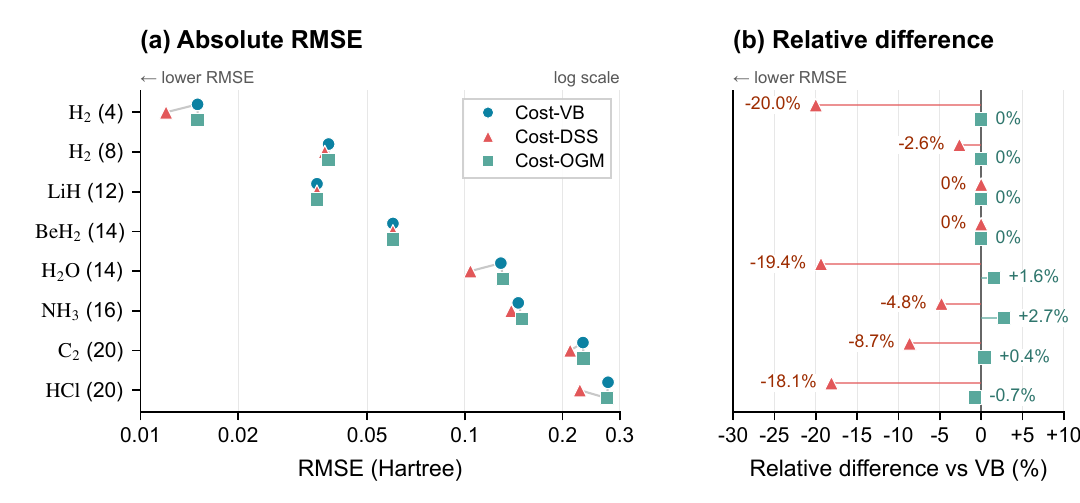}
    \caption{
    \textbf{Effect of the state-independent training proxy.} Ground-state energy RMSE for \model at $d_{\max}=2$ using VB, DSS, and OGM proxy costs.
    \textbf{(a)} Absolute RMSE in Hartree on a logarithmic horizontal scale, where lower values indicate more accurate energy estimates.
    \textbf{(b)} Percentage difference relative to the VB baseline results; negative values indicate lower RMSE. Blue filled circles, red filled triangles, and green filled squares denote the VB, DSS, and OGM proxy costs, respectively. The percentage annotations report the change obtained using the DSS proxy.
    }
    \label{fig:proxy-ablation}
\end{figure}

As shown in Fig.~\ref{fig:proxy-ablation}, the DSS confidence proxy matches or improves the VB proxy on all eight molecular systems. The largest RMSE reductions are $20.0\%$ for $\mathrm{H}_2(4)$, $19.4\%$ for $\mathrm{H}_2\mathrm{O}(14)$, and $18.1\%$ for $\mathrm{HCl}(20)$. It also reduces the RMSE by $8.7\%$ for $\mathrm{C}_2(20)$, $4.8\%$ for $\mathrm{NH}_3(16)$, and $2.6\%$ for $\mathrm{H}_2(8)$, while the results for LiH and $\mathrm{BeH}_2$ remain essentially unchanged. The OGM proxy does not yield a similarly consistent improvement. We therefore use the DSS confidence proxy for the main molecular results in Table~\ref{tab:molecular-rmse}.

These results show that the state-independent objective is a substantive component of the measurement-design framework rather than merely a training detail. They also show that the framework is modular in its training objective. The same \model pipeline was trained under all three proxies with the terminal reward as the only change, and each yielded complete measurement ensembles. Improving the correspondence between a classically computable proxy and the final finite-shot error remains an important direction for further development, and any improved proxy can be adopted directly as a training reward.

\subsection{Reusing the learned policy across a potential-energy surface}
\label{sec:pes-results}

A central advantage of learning a generative policy is that part of the measurement-design effort may be reused across related Hamiltonians. We test this possibility on an $\mathrm{H}_2\mathrm{O}$ potential-energy surface. We first train a \model policy at the equilibrium geometry, with an H--O--H angle of $104.5^\circ$ and equal O--H bond lengths of $0.9584$~\AA, and retain the learned network weights. These weights are then used to initialize training at a grid of symmetric geometries with common O--H bond lengths ranging from $0.8584$ to $1.0584$~\AA\ and H--O--H angles from $100^\circ$ to $109^\circ$. The standard baseline uses the same architecture, depth constraint, and training protocol but starts from random weights at every geometry. The two initialization schemes otherwise use identical measurement budgets, circuit constraints, proxy objectives, training budgets, and schedule selection procedures. To test whether the transferred measurement structure is tied to the training objective, we additionally perform a cross-proxy transfer: a single converged VB-proxy warm-start policy, obtained at the grid geometry nearest equilibrium ($104.5^\circ$, $0.9473$~\AA), is used to initialize training at every geometry under the DSS proxy, with all other settings unchanged. Full settings and the convergence criterion are given in Appendix~\ref{app:pes-details}.

\begin{figure}[htbp]
\centering
\includegraphics[width=0.47\linewidth]{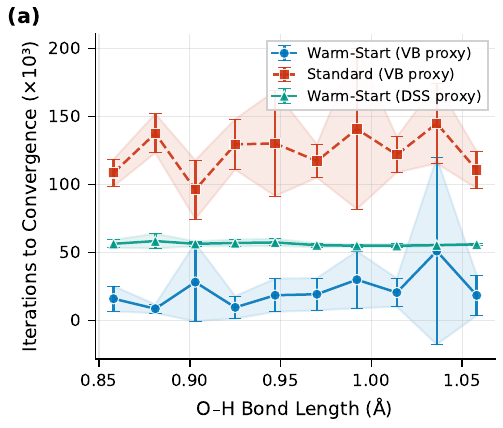}
\includegraphics[width=0.47\linewidth]{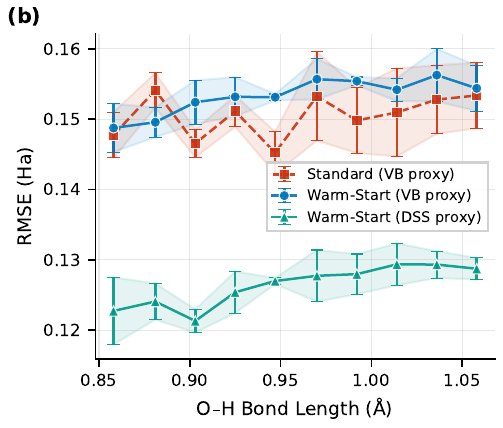}
\caption{\textbf{Reuse of the learned policy across an $\mathrm{H}_2\mathrm{O}$ potential energy surface.}
Number of gradient updates required to satisfy the convergence criterion (\textbf{a}) and final ground-state energy RMSE (\textbf{b}) as functions of the symmetric O--H bond length. At each bond length, the markers show averages over five H--O--H angles spanning $100^\circ$--$109^\circ$, while the error bars and shaded regions indicate the variation across angles. Warm-start training initializes \model from a policy trained at the reference geometry, whereas standard training uses random initialization at each geometry; both optimize the VB proxy. DSS-proxy runs instead optimize the DSS proxy, warm-started at every geometry from a single converged VB-proxy policy at the reference geometry.}
\label{fig:pes}
\end{figure}

Across the bond-length grid, warm-start initialization substantially reduces the number of iterations required for convergence. Convergence is determined from the persistent block-median criterion defined in Appendix~\ref{app:pes-details}, using the common reference horizon $T_{\mathrm{ref}}=5\times10^5$ gradient updates. Across bond lengths, the median convergence time over the five angles ranges from approximately $8\times10^3$ to $50\times10^3$ gradient updates for warm-start training, compared with approximately $95\times10^3$ to $145\times10^3$ updates for random initialization. At matched bond lengths, the ratio of the standard to warm start median convergence times ranges from approximately 3 to more than 10.

The cross-proxy transfer shows that the reused structure is not objective-specific. Initialized from the converged VB-proxy policy, DSS-proxy training reaches final RMSE values of $0.123$--$0.129$~Ha across the grid, below both VB-proxy arms ($0.145$--$0.157$~Ha) at every bond length, a reduction of $15$--$18\%$; the improvement holds at all 50 geometries individually. The first improvement of the sampled circuits under the new objective occurs after $55$--$58\times10^3$ updates, nearly independent of geometry.

This acceleration does not lead to a systematic loss in measurement quality. Between two VB proxy results, the final RMSE values remain comparable across the potential energy surface, with overlapping interquartile ranges at most bond lengths and no initialization consistently outperforming the other. The cross-proxy transfer, in contrast, improves measurement quality systematically: DSS-proxy training initialized from the converged VB-proxy policy reaches final RMSE values of $0.123$--$0.129$~Ha, below both VB-proxy results at every bond length. In both cases, the pretrained policy is not applied zero-shot; rather, it provides an effective initialization that transfers useful measurement structure between nearby Hamiltonians and across training objectives; it can substantially reduce the required retraining effort.

\subsection{Scaling to larger interacting fermionic systems}
\label{sec:hubbard-results}
The molecular benchmarks above rely on exact state-vector simulation and extend to 20 qubits. To test the practical computational scale of the measurement-design pipeline substantially beyond this regime, we consider compactly encoded spinless Hubbard models on $4\times4$ and $6\times6$ lattices, represented using 24 and 54 qubits, respectively~\cite{derby2021compact,dyrenkova2025scalable}. 

Reference ground states are obtained by two-site DMRG ~\cite{white1992density,schollwock2011density} on the corresponding $L^2$-qubit Jordan--Wigner Hamiltonians, using 16 and 36 qubits for the two lattice sizes. We do not construct an MPS directly on the 24 or 54 qubit compact registers. Instead, expectation values and covariances of encoded physical Pauli operators are evaluated by pulling those operators back through the compact-encoding isometry and contracting them with the Jordan--Wigner MPS.
Further details of the Hamiltonians, compact encoding, DMRG calculations, and oracle allocation are given in Appendix~\ref{app:hubbard-dmrg}.
Expectation values of observables on the compactly encoded registers are then evaluated through the compact-encoding pullback described in \ref{app:hubbard-mps}.

Under the same total measurement budget, we compare \model against an oracle Neyman allocation that distributes shots across a fixed set of QWC measurement groups in proportion to the square root of each group’s variance~\cite{wecker2015progress,crawford2021efficient,neyman1934two} (Appendix~\ref{app:hubbard_qwc}).
Both protocols use the same total measurement budget of $N=1000$. The \model schedules are generated with a maximum CNOT depth of $d_{\max}=2$ and are trained using the DSS proxy $C_{\mathrm{DSS}}$ in Eq.~\eqref{eq:cost_dss}, with $w_k=|c_k|$ and $\varepsilon=0.9$. As in the molecular experiments, the proxy depends only on the Hamiltonian coefficients and the state-independent Pauli hit counts of the generated ensemble.

The oracle baseline first constructs a fixed QWC partition and then allocates the measurement budget using the variances of the group energy contributions computed from the DMRG reference state. It therefore has access to target-state expectation values and covariances that would generally be unavailable before measurement. In contrast, neither the DMRG state nor any state-dependent moment is used to train \model or select its measurement schedule.

\begin{figure}[htbp]
    \centering
    \includegraphics[width=0.5\linewidth]{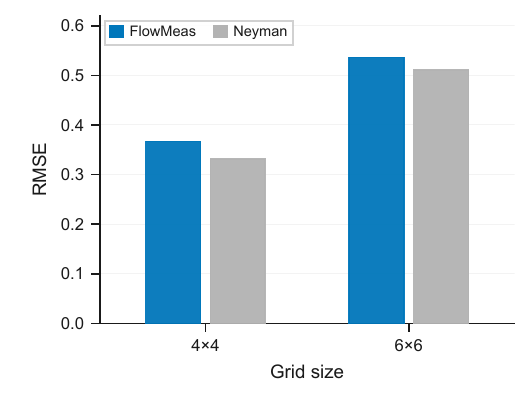}
    \caption{\textbf{Scaling to compactly encoded spinless Hubbard models.}
    Ground-state energy RMSE (arbitrary units) for $4\times4$ and $6\times6$ spinless Hubbard lattices, represented using 24 and 54 qubits, respectively. \model is compared with an oracle Neyman allocation over fixed QWC measurement groups under the same total measurement budget. The oracle uses group variances computed from the DMRG reference state, whereas \model uses a state-independent proxy. The reported RMSE values are computed from the exact conditional mean-squared error of each fixed schedule using MPS-derived first and second moments.
    }
    \label{fig:hubbard}
\end{figure}

As shown in Fig.~\ref{fig:hubbard}, the oracle Neyman allocation remains more accurate, as expected from its access to the target-state variances. Nevertheless, \model remains within the same error scale while using no state information during measurement design. More importantly, the direct generative pipeline remains tractable for the 54-qubit interacting model, far beyond the molecular systems accessible to exact state-vector simulation and the system sizes used in previous molecular measurement benchmarks.

Together with the 20-qubit molecular results, this experiment demonstrates that \model can construct state-independent measurement protocols at substantially larger scales than previously demonstrated for optimized molecular measurement design. It also illustrates the complementary roles of the two comparisons: the molecular benchmarks establish improvements in measurement accuracy, while the Hubbard models establish the computational reach of the framework.

\section{Discussion and outlook}
\label{sec:discussion}
Our central contribution is a unified generative framework for resource-constrained quantum measurement design. Rather than constructing measurement schedules instance by instance through grouping heuristics or sequential derandomization, \model formulates the design of a complete executable measurement protocol as learning a distribution over shallow Clifford circuits. In \model, hardware constraints define the support of a learned distribution, namely the allowed family of shallow Clifford circuits, while the statistical quality of a complete finite ensemble defines its reward, computed exactly by stabilizer propagation. A single formulation thus contains the standard measurement families as limiting cases and interpolates across the depth–sampling trade-off. At zero entangling depth the learned schedules already match or outperform leading product measurement methods; one or two entangling gate layers reduce the energy estimation error by up to $27\%$ relative to the strongest state-independent product baseline and match or improve DSS results; and the same pipeline extends to 20-qubit molecules and a compactly encoded 54-qubit fermionic model, beyond previously demonstrated scales. Because training produces a distribution rather than a single schedule, part of the search is amortized. Furthermore, a policy learned at one molecular geometry accelerates design across an entire potential energy surface by factors of 3 to more than 10. 

The potential energy surface experiment is a small instance of a much larger paradigm: amortizing measurement design across problems rather than within one. A policy conditioned explicitly on the target observables and their weights, the shot budget, and a machine readable description of the device could be pretrained on families of Hamiltonians, connectivity graphs, and calibration snapshots, then emit near-optimal measurement ensembles for unseen instances in a single forward pass. In this framework, measurement design stops being a per-molecule optimization and becomes a compilation pass in the quantum software stack: a learned measurement compiler, invoked as routinely as a transpiler, that translates a specification of what to estimate into hardware executable instructions for how to measure. The natural endpoint is a foundation model for quantum measurement, pretrained once across chemistry, materials, and device families, then specialized in seconds to each new molecule, budget, or calibration snapshot, making it practical to redesign the measurement layer as often as the device itself changes. 

Another possible direction would be to separate the learning into the two stages that made foundation models effective elsewhere in machine learning: pretraining with state-independent proxies and fine-tuning with state-dependent data. The proxies studied here are the natural pretraining signal. They are computable entirely classically, before the quantum device is ever queried, and are therefore available at the scale a pretraining corpus demands. Our comparison shows that the choice of proxy materially affects final accuracy, and because the objective enters \model only through the terminal reward, a sharper proxy can be adopted without changing anything else in the pipeline. The same modularity opens the fine-tuning stage: a policy pretrained on classical proxies can be adapted with small amounts of quantum data from the target experiment, with measured covariances replacing proxy assumptions~\cite{choi2022ghost,yen2023deterministic} and calibrated gate infidelities repricing each entangling layer. The next practical step is to demonstrate on hardware that the sampling advantage of shallow entangling ensembles survives realistic noise~\cite{hu2025demonstration}. Longer term, the measurement layer could be pretrained offline on classical proxies and fine-tuned in the loop on quantum data, running alongside the experiment rather than before it.

Most broadly, any procedure that converts quantum states into classical information, from correlation functions and reduced density matrices to Hamiltonian learning, device certification, and shadow tomography protocols~\cite{huang2023learning,huang2020predicting,carrasco2021verification,patel2025quantum,wu2025designing}, is specified by a weighted set of targets, a constrained protocol family, and an estimation risk, and can be cast in an analogous form. The same structure reappears at the logical level of early fault-tolerant machines, where measurements must be scheduled from fault-tolerant Clifford primitives under stringent routing and time constraints and every logical shot is expensive; designing such schedules is again discrete, compositional, resource-constrained generation. The ability of a generative policy to sample multiple high-reward solutions could also be useful for discovery, potentially exposing circuit structures not captured by existing analytic constructions. Beyond the specific protocol developed here, the broader design principle is that the interface between quantum devices and classical information need not be fixed by hand, but can itself become an object of learning and optimization.

\section{Methods}
\label{sec:methods}

\subsection{Measurement schedules and estimators}
\label{sec:measurement-estimator}

We use the $n$-qubit Hamiltonian introduced
in Eq.~\eqref{eq:ham}. A measurement schedule is represented as an ordered list $\mathbf U=(U_1,\ldots,U_N)$ of Clifford circuits, with one circuit assigned to each of the $N$ measurement shots. Operationally, this list is a multiset: repeated circuits correspond to repeated measurements in the same basis, and the ordering does not affect the resulting estimator.
Whenever circuit $U_j$ covers $P_k$, equivalently $\chi_{jk}=1$, the conjugated Pauli operator can be written as
\begin{equation}
    U_jP_kU_j^\dagger
    =
    s_{jk}\prod_{q\in T_{jk}}Z_q,
    \label{eq:pauli-diagonalization}
\end{equation}
where $s_{jk}\in\{-1,+1\}$and $T_{jk}\subseteq\{0,\ldots,n-1\}$. For each $P_k$, let $I_k(\mathbf U)=\{j\in\{1,\ldots,N\}:\chi_{jk}=1\}$ denote the set of shots that contribute to its estimator. By Eq.~\eqref{eq:hit-count}, we have $h_k(\mathbf U)=|I_k(\mathbf U)|$.

Let $b_j\in\{0,1\}^n$ denote the computational-basis outcome obtained after applying $U_j$. For each $j\in I_k(\mathbf U)$, the corresponding single-shot estimator is
\begin{equation}
    \xi_{jk}
    =
    s_{jk}
    \prod_{q\in T_{jk}}(-1)^{(b_j)_q}.
    \label{eq:single-shot-estimator}
\end{equation}
We estimate the Pauli expectation value using
\begin{equation}
    \widehat\mu_k(\mathbf U)
    =
    \begin{cases}
    \displaystyle
    \frac{1}{h_k(\mathbf U)}
    \sum_{j\in I_k(\mathbf U)}\xi_{jk},
    &
    h_k(\mathbf U)>0,
    \\[1.2ex]
    0,
    &
    h_k(\mathbf U)=0.
    \end{cases}
    \label{eq:pauli-estimator}
\end{equation}
The corresponding energy estimator is
\begin{equation}
    \widehat E(\mathbf U)
    =
    c_0+\sum_{k=1}^{M}c_k\widehat\mu_k(\mathbf U).
    \label{eq:energy-estimator}
\end{equation}

When every Pauli term is covered, $\widehat E$ is unbiased. Assigning the zero estimator to uncovered terms introduces a state-dependent bias. For a target state $\rho$, the conditional finite-shot risk is therefore the mean-squared error
\begin{equation}
    \mathcal R(\mathbf U;\rho)
    =
    \mathbb E_\rho
    \left[
        \left(
            \widehat E(\mathbf U)
            -
            \operatorname{Tr}(\rho H)
        \right)^2
        \,\middle|\,
        \mathbf U
    \right],
    \label{eq:true-risk}
\end{equation}
which includes both estimator variance and any bias caused by uncovered terms. This is the quantity ultimately reflected by the numerical RMSE, but it cannot generally be evaluated before information about $\rho$ is available.

\subsection{State-independent proxy costs}
\label{sec:cost-functions}

The true finite-shot risk depends on the unknown target state and, in general, on covariances between Pauli estimators, which state-dependent allocation schemes estimate explicitly~\cite{yen2023deterministic}. We therefore train \model using proxy costs that depend only on the Hamiltonian coefficients and the hit counts of the generated measurement ensemble. We compare three such objectives.

The first is a variance-plus-bias proxy,
\begin{equation}
C_{\mathrm{VB}}(\mathbf U)=
\sum_{k:h_k(\mathbf U)>0}\frac{c_k^2}{h_k(\mathbf U)}
+\left(\sum_{k:h_k(\mathbf U)=0}|c_k|\right)^2.
\label{eq:vb-cost}
\end{equation}
The first term is the diagonal Haar-averaged variance, up to the common factor $2^n/(2^n+1)$, while the second term is a state-independent upper bound on the squared bias produced by assigning the zero estimator to uncovered terms.

The second objective is exactly the weighted confidence cost used by DSS, specialized to deterministic measurement circuits~\cite{kirk_derandomized_2024}:
\begin{equation}
C_{\mathrm{DSS}}(\mathbf U)
=
\sum_{k=1}^{M}
w_k
\exp\!\left[
-\frac{\varepsilon^2}{2}h_k(\mathbf U)
\right].
\label{eq:cost_dss}
\end{equation}
Here $w_k$ is the importance weight of $P_k$; as in the DSS quantum chemistry experiments, we use $w_k=|c_k|$. The parameter $\varepsilon$ controls how strongly the cost penalizes small hit counts.

The third objective is the finite-schedule form of the OGM diagonal-variance cost with its finite-budget omission penalty~\cite{wu2023overlapped}:
\begin{equation}
C_{\mathrm{OGM}}(\mathbf U)
=
N\sum_{k:h_k(\mathbf U)>0}\frac{c_k^2}{h_k(\mathbf U)}
+N\sum_{k:h_k(\mathbf U)=0}c_k^2.
\label{eq:ogm-cost}
\end{equation}
Multiplication by the common positive factor $N$ does not change the ordering of measurement ensembles, although it must be accounted for consistently in reward normalization. Derivations and comparisons of the three objectives are given in Appendix~\ref{app:proxy-costs}.

\subsection{Direct Clifford circuit generator}
\label{sec:circuit-generator}

Each partial circuit is represented by a Clifford tableau~\cite{gottesman1998heisenberg,aaronson2004improved} together with the construction metadata required to enforce the gate ordering and depth constraints. For the purpose of determining Pauli coverage, signs can be tracked separately from the binary symplectic action. A term is covered precisely when its propagated Pauli has no $X$ component after conjugation by the completed circuit.

\begin{figure}[htbp]
    \centering
    \includegraphics[width=0.78\linewidth]{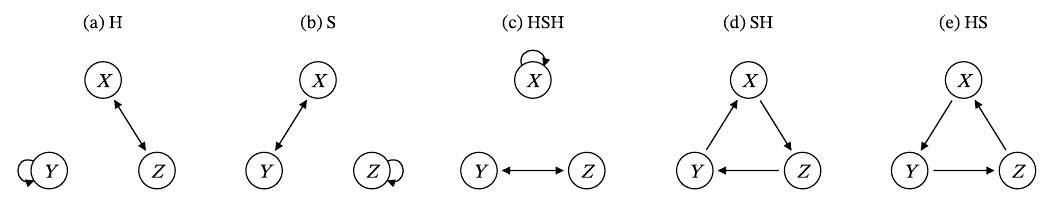}
    \caption{\textbf{Single-qubit Clifford representatives.}
    Conjugation by $H$, $S$, $HSH$, $SH$, and $HS$ generates the five nonidentity permutations of the Pauli axes, up to signs. Together with the identity, these actions represent the six elements of the single-qubit Clifford group modulo Pauli operators and global phase that are relevant to Pauli coverage.}
    \label{fig:clifford}
\end{figure}

Ignoring Pauli signs, the single-qubit Clifford group induces the six permutations of the Pauli axes (Fig.~\ref{fig:clifford}). We represent these six classes by the identity and the five nontrivial gates
\begin{equation}
    H,\qquad S,\qquad HS,\qquad SH,\qquad HSH.
\end{equation}
Here the equivalence relation removes Pauli factors and global phases, which do not affect whether a Pauli string is mapped into $\pm\{I,Z\}^{\otimes n}$.
The canonical circuit representation alternates local-Clifford blocks and entangling CNOT layers,
\begin{equation}
    U
    =
    L_D E_D L_{D-1} E_{D-1}\cdots L_1 E_1 L_0,
    \label{eq:canonical-circuit-form}
\end{equation}
where $L_\ell=\bigotimes_{q=0}^{n-1}C_{\ell,q}$, with $C_{\ell,q}\in \{I,H,S,HS,SH,HSH\}$, and each $E_\ell$ is a layer of directed nearest-neighbor CNOT gates with pairwise-disjoint supports. The number of CNOT layers satisfies $0\leq D\leq d_{\max}$.

Equation~\eqref{eq:canonical-circuit-form} is a normal form for the generated circuit rather than the order in which actions are sampled. In the implementation, single-qubit and CNOT actions may be interleaved. Within a given CNOT layer, however, all CNOT supports are disjoint. Consequently, a single-qubit action interleaved among CNOTs in the same layer can always be commuted to the appropriate neighboring local block: if its qubit has already been used in that layer, it commutes with all subsequent CNOTs in the layer; if it has not yet been used, it commutes with all preceding CNOTs. Together with the single-qubit compression rule described below, every generated sequence therefore admits the normal form above.

The implementation nevertheless retains the original serialized action order exactly as sampled; no gate reordering is performed during generation or evaluation. The identity choices in each $L_\ell$ are implicit and are represented by the absence of a local action.

The base action set is
\begin{equation}
\begin{split}
    \mathcal A_n
    ={}&
    \{
        H_q,S_q,HS_q,SH_q,HSH_q:
        0\leq q<n
    \}
    \\
    &\cup
    \{
        \operatorname{CNOT}_{q,q+1},
        \operatorname{CNOT}_{q+1,q}:
        0\leq q<n-1
    \}
    \cup\{\bot\}.
    \label{eq:action-set}
\end{split}
\end{equation}
where $\bot$ is the terminal action. Actions are applied directly to the serialized partial circuit. CNOT actions additionally update the current greedy CNOT-layer metadata defined below.

An action mask restricts $\mathcal A_n$ according to the full state of the current partial circuit. The implementation does not use alternating sequence of completed single-qubit blocks and CNOT layers. Single-qubit and CNOT actions may be
interleaved arbitrarily whenever they satisfy the masks described below. The serialized gate order is exactly the sampled action order. No post-generation reordering of gates acting on different qubits is applied.

More precisely, the full construction state at step $t$ can be written as
\begin{equation}
    s_t
    =
    \left(
        T_t,
        a_{1:t},
        d_t,
        \mathcal Q_t,
        \mathbf r_t,
        \boldsymbol{\ell}_t,
        \zeta_t
    \right),
    \label{eq:construction-state}
\end{equation}
where $T_t$ is the sign-free binary symplectic matrix of the tableau, $a_{1:t}$ is the serialized action history, $d_t$ is the number of opened CNOT layers, $\mathcal Q_t\subseteq\{0,\ldots,n-1\}$ is the set of qubits occupied in the current CNOT layer, $\mathbf r_t$ records the most recent single-qubit Clifford action on each qubit since the last CNOT applied to that qubit, $\boldsymbol{\ell}_t$ records the last action step using each qubit, and $\zeta_t$ indicates whether the circuit has terminated. At initialization, $T_0=I_{2n}$, $a_{1:0}=\varnothing$, $d_0=0$, $\mathcal Q_0=\varnothing$, $\mathbf r_0(q)=\varnothing$ and $\boldsymbol{\ell}_0(q)=-1$ for every qubit $q$, and $\zeta_0=\mathrm{false}$. These metadata determine the valid-action set
\begin{equation}
    \mathcal A(s_t)\subseteq\mathcal A_n.
\end{equation}

The policy observes only the binary symplectic component of $s_t$. Specifically, its observation is the vectorized symplectic tableau,
\begin{equation}
o_t=\operatorname{vec}(T_t),
\label{eq}
\end{equation}
where $\operatorname{vec}$ denotes vectorization of the $2n\times2n$ binary matrix. The observation space is
$\mathcal{O}=\left\{\operatorname{vec}(T) \mid T \in \operatorname{Sp}\left(2 n, \mathbb{F}_2\right)\right\} \subseteq \mathbb{F}_2^{4 n^2}$.

Let $f_\theta:\mathcal{O} \rightarrow\mathbb{R}^{|\mathcal{A}_n|}$ denote the neural policy network where , which maps the tableau observation $o_t$ to one unnormalized logit for each action in the base action set $\mathcal A_n$. The environment then masks actions that are invalid in the full construction state $s_t$.
Equivalently, the masked forward distribution is

\begin{equation}
    P_F^\theta(a\mid s_t)
    =
    \frac{
        \exp\!\left(f_\theta(o_t)_a\right)
        \mathbf{1}\!\left[a\in\mathcal A(s_t)\right]
    }{
        \displaystyle
        \sum_{a'\in\mathcal A_n}
        \exp\!\left(f_\theta(o_t)_{a'}\right)
        \mathbf{1}\!\left[a'\in\mathcal A(s_t)\right]
    }.
    \label{eq:masked-policy}
\end{equation}

Thus, two partial constructions may have the same tableau observation but different construction metadata and therefore different valid-action masks. They are distinct construction states even though the neural policy receives the same tableau input. 

The single-qubit action mask implements local-Clifford compression on each qubit. Let $r_t(q)=\varnothing$ indicate that no single-qubit Clifford has been applied to qubit $q$ since the most recent CNOT applied to that qubit. A single-qubit action $C_q$ is valid only when $r_t(q)=\varnothing$. Applying it sets $r_{t+1}(q)=C$. Applying a CNOT to qubits $q$ and $q'$ resets both $r_{t+1}(q)$ and $r_{t+1}(q')$ to $\varnothing$. Actions on other qubits do not alter these values. Consequently, between two consecutive CNOTs involving a given qubit, there can be either zero or one nonidentity single-qubit Clifford on that qubit. In particular, this rule does not forbid cascading CNOTs.
This is not a restriction on the represented sign-free local Clifford action, because any product of single-qubit Clifford gates in such an interval can be composed and replaced by one of the six classes $\{I,H,S,HS,SH,HSH\}$.
The identity class is represented by taking no local action. Importantly, this compression is enforced independently on each qubit. 
A single-qubit action on one qubit does not prevent a subsequent single-qubit action on another qubit. For example, $[ H_0, \operatorname{CNOT}_{0,1}, H_2]$ is a valid action sequence. 
The action $H_2$ updates the tableau and the local-action record for qubit $2$, but it does not change the current CNOT-layer metadata $d_t$ or $\mathcal Q_t$.

CNOT depth is assigned online using a greedy layer-packing rule. Consider a valid directed nearest-neighbor CNOT acting on the unordered support $e=\{q,q'\}$.
Its layer update is
\begin{equation}
    (d_{t+1},\mathcal Q_{t+1})
    =
    \begin{cases}
        \left(1,e\right),
        &
        d_t=0,
        \\[3pt]
        \left(d_t,\mathcal Q_t\cup e\right),
        &
        d_t>0
        \ \text{and}\
        e\cap\mathcal Q_t=\varnothing,
        \\[3pt]
        \left(d_t+1,e\right),
        &
        e\cap\mathcal Q_t\neq\varnothing.
    \end{cases}
    \label{eq:cnot-layer-update}
\end{equation}
Thus, the first CNOT opens the first CNOT layer. A subsequent CNOT whose support is disjoint from all CNOTs already assigned to the current layer is packed into that same layer. If either of its qubits is already occupied in the current layer, it opens a new layer and resets the current-layer occupancy to its own two-qubit support. A CNOT requiring a new layer is masked when $d_t=d_{\max}$, and all CNOT actions are masked when $d_{\max}=0$. Single-qubit actions do not change either $d_t$ or $\mathcal Q_t$. Therefore, the current CNOT layer remains open across intervening single-qubit actions.

The action mask additionally removes a redundant repeated-CNOT pattern. A candidate CNOT on the pair $\{q,q'\}$ is masked as $\ell_t(q)=\ell_t(q')=u\geq 0$, and the action at step $u$ was itself a CNOT on the same unordered pair $\{q,q'\}$. Both control-target orientations are masked in this case. An intervening action on either $q$ or $q'$ changes its last-use step and therefore lifts this mask, whereas actions involving only unrelated qubits do not. Together with the single-qubit compression rule, this removes simple local redundancies without identifying all action sequences that implement the same Clifford unitary.

The resulting construction graph is nevertheless finite and acyclic. Every nonterminal transition appends exactly one action to the history $a_{1:t}$, and no forward transition removes or reorders an earlier action. The action-history length therefore provides a strictly increasing rank for every forward edge. Moreover, each qubit can receive at most one single-qubit action before the first CNOT applied to it and at most one additional single-qubit action after each CNOT layer that touches it. Since a qubit occurs in at most one CNOT per layer, the number of nonidentity single-qubit actions is bounded by $n(d_{\max}+1)$.
Each CNOT layer contains at most $\lfloor n/2\rfloor$ pairwise disjoint CNOTs, so the number of CNOT actions is bounded by
$d_{\max}\lfloor n/2\rfloor$.
Hence every circuit terminates after at most
\begin{equation}
    n(d_{\max}+1)
    +
    d_{\max} \lfloor \frac{n}{2}\rfloor
    +1
    \label{eq:max-trajectory-length}
\end{equation}
actions, including the stop action. The stop action $\bot$ maps the current construction to a terminal sink. Circuits may terminate before exhausting the CNOT-depth budget. Empty circuits are allowed and correspond to direct computational-basis measurement. Different measurement slots may also terminate at identical reduced circuit descriptions, so repeated circuits in the measurement ensemble are allowed.

\subsection{Ensemble construction and GFlowNet objective}
\label{sec:gflownet-objective}

A standard GFlowNet is defined on a directed acyclic graph of partially constructed objects~\cite{bengio2021flow,malkin2022trajectory}. Starting from a source state, a forward policy generates a trajectory terminating at an object $x$. Given a positive reward $R(x)$ and a backward policy over valid reverse transitions, trajectory balance requires the forward trajectory flow to match the reward-weighted backward trajectory flow~\cite{malkin2022trajectory}. Alternative credit-assignment schemes based on partial trajectories are also available~\cite{madan2023subtb}; we use the full-trajectory objective throughout.

For \model, the terminal object is a complete measurement schedule. At construction step $t$, let $S_t=(s_{1,t},\ldots,s_{N,t})$ denote the collection of partial circuit states, and let $\mathcal I_t$ be the set of circuit slots that have not yet terminated. A shared circuit-level policy samples one action for each active slot. The joint forward transition probability is
\begin{equation}
    P_F^\theta(A_{t+1}\mid S_t)
    =
    \prod_{j\in\mathcal I_t}
    P_F^\theta(a_{j,t+1}\mid s_{j,t}),
    \label{eq:factorized-forward}
\end{equation}
where the ensemble actions $A_{t+1} = \{a_{j,t+1}:j\in\mathcal I_t\}$.
Slots that sample $\bot$ become terminal and remain unchanged thereafter. The construction ends when all slots have terminated. A complete ensemble trajectory is
\begin{equation}
    \boldsymbol\tau
    =
    (S_0,A_1,S_1,\ldots,A_T,S_T=\mathbf U).
\end{equation}
Its forward probability is
\begin{equation}
    P_F^\theta(\boldsymbol\tau)
    =
    \prod_{t=0}^{T-1}
    \prod_{j\in\mathcal I_t}
    P_F^\theta(a_{j,t+1}\mid s_{j,t}).
    \label{eq:ensemble-path-probability}
\end{equation}
The terminal reward $R(\mathbf U)$ is evaluated only after the complete schedule has been generated, because the usefulness of a circuit depends on the coverage supplied by the other circuits.

Let $P_B(\boldsymbol\tau)$ denote the probability of the corresponding reverse trajectory under the backward policy, and let $Z_\theta$ be the learned total flow. The ensemble trajectory-balance residual is
\begin{equation}
    \Delta_{\mathrm{ens}}(\boldsymbol\tau)
    =
    \log Z_\theta
    +
    \log P_F^\theta(\boldsymbol\tau)
    -
    \log R(\mathbf U)
    -
    \log P_B(\boldsymbol\tau).
    \label{eq:ensemble-tb}
\end{equation}
Training minimizes
\begin{equation}
    \mathcal L_{\mathrm{TB}}
    =
    \mathbb E_{\boldsymbol\tau\sim q_{\mathrm{train}}}
    \left[
    \Delta_{\mathrm{ens}}(\boldsymbol\tau)^2
    \right],
    \label{eq:tb-loss}
\end{equation}
where $q_{\mathrm{train}}$ denotes the trajectory-sampling distribution used to form training batches. In our implementation, training uses primarily masked on-policy trajectories, supplemented periodically by replay of retained high-reward trajectories.
The backward policy is fixed and uniform over the valid parents of each non-source state.

The forward policy uses a simple parameterization; all slots share one circuit-level policy, which acts on a single partial circuit without conditioning on the other slots or on the current ensemble hit counts. This keeps the policy input size independent of the number of circuits and matches the permutation symmetry of the ensemble reward. Conditional on the learned parameters, the $N$ circuits are therefore i.i.d. draws from a common single-circuit marginal, and \model learns a distribution over ensembles within the product family $p_\theta(U_1,\ldots,U_N)=\prod_{j=1}^N p_\theta(U_j)$.

Training is fully ensemble-level. The reward $R(\mathbf U)$ and the residual in Eq.~\eqref{eq:ensemble-tb} are evaluated on complete schedules, so the ensemble objective shapes the shared marginal even though the circuits remain independent during generation. Because $R(\mathbf U)$ does not factorize across circuits, no product distribution can be exactly proportional to it; trajectory balance therefore serves as a learning objective within this family rather than an exact sampler, and it biases the marginal circuit generator toward distributions whose finite samples produce low-cost schedules.

\subsection{Policy training and schedule selection}
\label{sec:training-protocol}

As described in Sec.~\ref{sec:gflownet-objective}, the shared forward policy acts on the Clifford tableau
representation of a single partial circuit. It does not condition
on the Hamiltonian coefficients, target Pauli strings, current
hit counts, or the states of the other circuit slots; these
quantities influence learning only through the terminal ensemble
reward. A separate policy is trained for each Hamiltonian and
CNOT-depth setting. Transfer between related Hamiltonians is
implemented by reusing the learned parameters as an
initialization.

During training, the lowest-proxy-cost schedules encountered are
retained. The final measurement protocol is the best retained
schedule according to the same state-independent proxy used for
training and is fixed before measurement outcomes are generated.
The target-state RMSE is used only for retrospective evaluation,
not for schedule selection. The neural-network architecture and
complete training details are given in
Appendix~\ref{app:numerical-details}.

\section{Data Availability}

The molecular Hamiltonians used in this work were obtained from the public \texttt{variances} collection~\cite{hadfieldVariances}. All remaining numerical data were generated using publicly available software packages and the numerical procedures described in this manuscript. The code and data are available upon reasonable request. 

\section{Acknowledgements}

J.D. thanks Ewan Murphy and the Mila IDT team for helpful discussions on GPU implementations of stabilizer tableau propagation. O.N.L. acknowledges the financial support of the Ministère de l’\'{E}conomie, de l’Innovation et de l’\'{E}nergie du Québec (MEIE). C.Z. acknowledges support from the Natural Sciences and Engineering Research Council of Canada (NSERC) through Discovery Grant RGPIN-2026-06304; from the Faculté des sciences and the Institut quantique at Université de Sherbrooke; and from the Institut transdisciplinaire d’information quantique (INTRIQ), a strategic cluster funded by the Fonds de recherche du Québec – Nature et technologies.
G.R acknowledges support from the CIFAR AI Chair program, NSERC through Discovery Grant RGPIN-2026-06522 and IVADO's Exploratory Projects Program.
This research was enabled in part by support provided by Calcul Québec and the Digital Research Alliance of Canada.

GPT-5.6 Sol (OpenAI) and Claude 4.8 Opus (Anthropic) were used to assist with language editing, including grammar, spelling, and wording; clarifying certain technical concepts; and the implementation of standard methods. All AI-assisted outputs were critically reviewed, verified where appropriate, and revised by the authors. The authors take full responsibility for the scientific content, accuracy, and integrity of the manuscript.

\printbibliography
\newpage
\appendix

\section{State-independent proxy objectives}
\label{app:proxy-costs}

This appendix derives the three proxy costs introduced in Sec.~\ref{sec:cost-functions}. Throughout, we use the Hamiltonian notation of Eq.~\eqref{eq:ham}. We let $\mathbf U=(U_1,\ldots,U_N)$ denote a deterministic measurement schedule and use the hit count $h_k(\mathbf U)$ defined in Eq.~\eqref{eq:hit-count}.

\subsection{Variance-plus-bias proxy}
\label{app:vb-proxy}

For a fully covered schedule, the conditional variance of the
energy estimator defined in Eq.~\eqref{eq:energy-estimator} can be
written as
\begin{equation}
\operatorname{Var}_{\rho}
\!\left(\widehat E\mid\mathbf U\right)
=
\sum_{k=1}^{M}
c_k^2
\operatorname{Var}_{\rho}
\!\left(\widehat\mu_k\mid\mathbf U\right)
+
\sum_{k\neq\ell}
c_kc_\ell
\operatorname{Cov}_{\rho}
\!\left(
\widehat\mu_k,\widehat\mu_\ell
\mid\mathbf U
\right).
\label{eq:app-energy-variance}
\end{equation}

For a nonidentity Pauli operator $P_k$ estimated from
$h_k$ independent hits,
\begin{equation}
\operatorname{Var}_{\rho}
\!\left(\widehat\mu_k\mid\mathbf U\right)
=
\frac{1-\mu_k^2}{h_k},
\qquad
\mu_k=\operatorname{Tr}(\rho P_k).
\label{eq:app-pauli-variance}
\end{equation}

Let $D=2^n$. Under the Haar measure on pure states,
\begin{equation}
\mathbb E_{\mathrm{Haar}}[\mu_k^2]=\frac{1}{D+1},
\qquad
\mathbb E_{\mathrm{Haar}}[\mu_k\mu_\ell]=0
\quad(k\neq\ell),
\label{eq:app-haar-pauli-moments}
\end{equation}
for distinct nonidentity Pauli strings. The Haar-averaged covariance contributions in Eq.~\eqref{eq:app-energy-variance} therefore vanish, and the covered-term variance is
\begin{equation}
\mathbb E_{\mathrm{Haar}}
\left[
\operatorname{Var}_{\rho}(\widehat E\mid\mathbf U)
\right]
=
\frac{D}{D+1}
\sum_{k:h_k>0}\frac{c_k^2}{h_k}.
\label{eq:app-haar-variance}
\end{equation}
The prefactor $D/(D+1)$ is independent of the measurement schedule and is omitted from the proxy.

When $h_k=0$, we assign the zero estimator to $P_k$. The resulting energy bias is
\begin{equation}
B(\mathbf U;\rho)
=
-\sum_{k:h_k=0}c_k\operatorname{Tr}(\rho P_k).
\label{eq:app-uncovered-bias}
\end{equation}
Using $|\operatorname{Tr}(\rho P_k)|\leq1$ and the triangle inequality gives
\begin{equation}
|B(\mathbf U;\rho)|^2
\leq
\left(\sum_{k:h_k=0}|c_k|\right)^2.
\label{eq:app-bias-bound}
\end{equation}
Combining the schedule-dependent part of Eq.~\eqref{eq:app-haar-variance} with Eq.~\eqref{eq:app-bias-bound} gives Eq.~\eqref{eq:vb-cost}. This is a hybrid state-independent surrogate: it combines a Haar-averaged variance for covered terms with a worst-case bias bound for uncovered terms. 

\subsection{DSS confidence cost}
\label{app:dss-proxy}

DSS associates each target Pauli string $P_k$ with an
importance weight $w_k$ and defines the cost of a collection
of possibly randomized circuit ensembles
$\{\mathcal U_j\}_{j=1}^{N}$ as~\cite{kirk_derandomized_2024}
\begin{equation}
C_{\mathrm{DSS}}
\!\left(\{\mathcal U_j\}_{j=1}^{N}\right)
=
\sum_{k=1}^{M}w_k
\prod_{j=1}^{N}
\exp\!\left[
-\frac{\varepsilon^2}{2}p_j(P_k)
\right],
\label{eq:app-dss-general}
\end{equation}
where
\begin{equation}
p_j(P_k)
=
\frac{1}{2^n}
\mathbb E_{U\sim\mathcal U_j}
\sum_{b\in\{0,1\}^n}
\left|
\langle b|UP_kU^\dagger|b\rangle
\right|^2
\label{eq:app-dss-pauli-weight}
\end{equation}
is the probability that a circuit drawn from
$\mathcal U_j$ diagonalizes $P_k$ in the computational
basis.

For the deterministic Clifford circuits generated by
\model, each $\mathcal U_j$ is a point mass at $U_j$, so
$p_j(P_k)\in\{0,1\}$ and
\begin{equation}
\sum_{j=1}^{N}p_j(P_k)
=
h_k(\mathbf U).
\label{eq:app-dss-hit-reduction}
\end{equation}
Therefore,
\begin{equation}
C_{\mathrm{DSS}}(\mathbf U)
=
\sum_{k=1}^{M}w_k
\exp\!\left[
-\frac{\varepsilon^2}{2}h_k(\mathbf U)
\right],
\label{eq:app-dss-deterministic}
\end{equation}
which is Eq.~\eqref{eq:cost_dss}. 
Here $w_k$ is the importance weight of $P_k$; as in the quantum chemistry experiments, we use $w_k=|c_k|$. The parameter $\varepsilon$ controls how strongly the cost penalizes small hit counts. Unless otherwise stated, we use $\varepsilon=0.9$ throughout all experiments employing the DSS proxy.

The two-sided Hoeffding confidence bound contains an
additional overall factor of two. We follow the main-text
DSS cost convention, which omits this constant because it
does not affect the optimization. The parameter
$\varepsilon$ is a cost-function hyperparameter and should
not be confused with the observed energy estimation error.

DSS evaluates the probabilities $p_j(P_k)$ for partially
randomized circuits during sequential gate fixing. In
\model, all generated circuits are deterministic, and the
same objective is evaluated directly from their exact hit
counts.

\subsection{OGM diagonal-variance objective}
\label{app:ogm-proxy}

OGM is formulated for a probability distribution $K$ over
product-Pauli measurement bases $B$. For a target Pauli
operator $P_k$, it defines the coverage probability
\begin{equation}
\chi_K(P_k)
=
\sum_{B:\,P_k\preceq B} K(B),
\label{eq:app-ogm-coverage}
\end{equation}
where $P_k\preceq B$ means that the basis $B$ measures
$P_k$. Its state-independent diagonal-variance objective with
a finite-budget omission penalty is~\cite{wu2023overlapped}
\begin{equation}
C_{\mathrm{OGM}}(K)
=
\sum_{k\in\mathcal S(K)}
\frac{c_k^2}{\chi_K(P_k)}
+
N\sum_{k\notin\mathcal S(K)}c_k^2,
\label{eq:app-ogm-general}
\end{equation}
where $\mathcal S(K)$ is the set of Pauli strings measured
with nonzero probability and $N$ is the total measurement
budget.

For the shallow Clifford measurement settings generated by
\model, we use the same coverage-probability form, replacing
the product-Pauli bases by the actual deterministic Clifford
circuits. A schedule
$\mathbf U=(U_1,\ldots,U_N)$ induces the empirical
distribution
\begin{equation}
K_{\mathbf U}(U)
=
\frac{1}{N}
\sum_{j=1}^{N}
\mathbf 1[U_j=U].
\label{eq:app-ogm-empirical}
\end{equation}
The corresponding probability that a randomly selected
circuit from the schedule diagonalizes $P_k$ is
\begin{equation}
\chi_{K_{\mathbf U}}(P_k)
=
\sum_U
K_{\mathbf U}(U)\,
\mathbf 1
\!\left[
U P_k U^\dagger
\in \pm\{I,Z\}^{\otimes n}
\right]
=
\frac{h_k(\mathbf U)}{N}.
\label{eq:app-ogm-hit-prob}
\end{equation}
Substituting this empirical coverage probability into
Eq.~\eqref{eq:app-ogm-general} gives
\begin{equation}
C_{\mathrm{OGM}}(\mathbf U)
=
N\sum_{k:h_k>0}\frac{c_k^2}{h_k}
+
N\sum_{k:h_k=0}c_k^2,
\label{eq:app-ogm-deterministic}
\end{equation}
which is Eq.~\eqref{eq:ogm-cost}. Unlike the squared joint
omission penalty in Eq.~\eqref{eq:vb-cost}, the OGM omission
penalty is separable across uncovered Pauli terms.

\subsection{Reward normalization}
\label{app:reward-normalization}

To keep the proxy values numerically stable across Hamiltonians with different coefficient scales, the Pauli coefficients are normalized before the proxy is evaluated. 
For each Hamiltonian, we define $c_{\max}=\max_{1\leq k\leq M}|c_k|$, the normalized coefficients $\overline c_k=c_k/c_{\max}$, and the corresponding normalized DSS weights $\overline w_k=|\overline c_k|=|c_k|/c_{\max}$.
The maximum is taken only over the nonidentity Pauli terms; the identity coefficient $c_0$ is excluded because it does not enter the measurement-design objective. This normalization is performed separately for each Hamiltonian and preserves the relative magnitudes of all nonidentity Pauli coefficients. 
Thus, VB and OGM use the normalized coefficients $\overline c_k$, and DSS uses the normalized weights $\overline w_k$.

For any proxy $C$, the terminal reward is
\begin{equation}
\log R(\mathbf U)
=
\alpha-\beta\widetilde C(\mathbf U),
\label{eq:app-log-reward}
\end{equation}
where $\widetilde C$ is the normalized cost used in the implementation. Before evaluating a proxy, the nonidentity Pauli coefficients are divided by their maximum absolute value. The reward parameters used in the reported experiments are $\alpha=10$ and $\beta=100$.

\section{Numerical and implementation details}
\label{app:numerical-details}

All reported protocols first select a fixed measurement schedule and then evaluate that schedule using independent measurement outcomes from a common reference state. No target-state expectation values or covariances are used to train \model.

\subsection{Implementation and training environment}
\label{app:implementation}

The implementation uses Python, PyTorch~\cite{paszke2019pytorch}, CuPy~\cite{okuta2017cupy}, and custom CUDA kernels. Clifford propagation, Pauli diagonalization checks, hit-matrix construction, and proxy evaluation are performed on the GPU. Unless stated otherwise, each training run uses one NVIDIA H100 GPU in a node with an AMD EPYC 9454 CPU and CUDA 13.0. A separate policy is trained for each Hamiltonian and CNOT-depth setting. The terminal measurement schedule contains $N=1000$ circuit entries for the molecular experiments, equal to the measurement budget used in the subsequent energy-estimation trials.

\begin{table}[H]
\centering
\small
\caption{Training and implementation hyperparameters.}
\label{tab:training-hyperparameters}
\begin{tabular}{ll}
\toprule
Quantity & Value\\
\midrule
Optimizer & Adam\\
Policy learning rate & $10^{-3}$\\
Log-flow learning rate & $10^{-2}$\\
Adam parameters & $\beta_1=0.9$, $\beta_2=0.999$, $\epsilon_{\rm Adam}=10^{-8}$\\
Weight decay & 0\\
Batch size & 16 ensembles; 8 for the 54-qubit instance\\
Maximum training length & $2\times10^6$ gradient updates\\
Gradient clipping & global norm clipped at 10\\
Learning-rate schedule & none\\
Exploration & masked on-policy sampling\\
Replay & ten highest-reward trajectories every 100 on-policy updates\\
\bottomrule
\end{tabular}
\end{table}

\subsection{GFlowNet training and schedule selection}
\label{app:training-details}

Training minimizes the ensemble trajectory-balance loss defined in Eq.~\eqref{eq:tb-loss}. The scalar $\log Z_\theta$ is learned jointly with the policy and initialized to zero. It is optimized with the same Adam optimizer as the policy parameters, using the log-flow learning rate listed in Table~\ref{tab:training-hyperparameters}. The backward policy is fixed and uniform over all valid parent states. Fresh trajectories are sampled on policy. A replay buffer stores the ten highest-reward terminal trajectories observed so far and replays them every 100 on-policy updates.

During training, the lowest-cost terminal ensemble encountered is retained. The final measurement schedule is selected according to the same state-independent proxy used for training and is fixed before quantum measurement outcomes are sampled. Ground-state RMSE and MAE are used only for retrospective evaluation.

\subsection{Neural-network architecture}
\label{app:policy-architecture}

For $n$ qubits, the action set in
Eq.~\eqref{eq:action-set} has size
\begin{equation}
    |\mathcal A_n|
    =
    5n+2(n-1)+1
    =
    7n-1.
    \label{eq:app-action-count}
\end{equation}
The shared forward policy is a multilayer perceptron whose input
is the flattened Clifford tableau (4n$^2$, 1) of a single partial circuit. It
has three hidden layers of width 1024, each using LayerNorm
followed by a leaky-ReLU nonlinearity. A linear output head
produces logits for the $7n-1$ actions, and invalid actions are
assigned logit $-\infty$ before the softmax. The architecture is
fixed across all systems and depth settings.

The policy input does not include Hamiltonian coefficients,
target Pauli strings, ensemble hit counts, or information about
the other circuit slots.

\subsection{Molecular Hamiltonians}
\label{app:molecular-hamiltonians}

The molecular benchmarks use Jordan--Wigner Hamiltonians from the public \texttt{variances} collection~\cite{hadfieldVariances}. Unless otherwise stated, the \texttt{jw.txt} file is used. The identity coefficient is excluded from the measurement-design objective and restored in the final energy estimate. Hamiltonian coefficients are expressed in Hartree. The systems and geometries are summarized in Table~\ref{tab:molecular-hamiltonians}.

\begin{table}[H]
\centering
\small
\caption{Molecular benchmark Hamiltonians. Distances and Cartesian coordinates are in \AA.}
\label{tab:molecular-hamiltonians}
\begin{tabular}{lcccl}
\toprule
System & Qubits & Basis & Mapping & Geometry\\
\midrule
$\mathrm{H}_2$ & 4 & STO-3G & JW & bond length 0.735\\
$\mathrm{H}_2$ & 8 & 6-31G & JW & bond length 0.7462\\
$\mathrm{LiH}$ & 12 & STO-3G & JW & bond length 1.548\\
$\mathrm{BeH}_2$ & 14 & STO-3G & JW & linear; Be--H distance 1.3038\\
$\mathrm{H}_2\mathrm{O}$ & 14 & STO-3G & JW & O $(0,0,0.137)$; H $(0,\pm0.769,-0.546)$\\
$\mathrm{NH}_3$ & 16 & STO-3G & JW & N $(0,0,0.149)$; H $(0,0.947,-0.348)$, $(\pm0.821,-0.474,-0.348)$\\
$\mathrm{C}_2$ & 20 & STO-3G & JW & bond length 1.2691\\
$\mathrm{HCl}$ & 20 & STO-3G & JW & bond length 1.343\\
\bottomrule
\end{tabular}
\end{table}

Reference ground states for these systems are obtained by exact diagonalization of the full qubit Hamiltonian.

\subsection{Potential energy surface training protocol}
\label{app:pes-details}

The potential energy surface experiment uses ten symmetric O--H bond lengths uniformly spanning $0.8584$--$1.0584$~\AA\ and five H--O--H angles spanning $100^\circ$--$109^\circ$, giving 50 target geometries. At each geometry, the warm-start run is initialized from the policy trained at the reference geometry, whereas the standard run uses randomly initialized policy weights. Both runs use the same architecture, measurement budget $N=1000$, maximum CNOT depth $d_{\max}=2$, VB proxy $C_{\mathrm{VB}}$, training budget, and schedule-selection procedure.

For each run, convergence is determined from the normalized proxy-cost trace $\widetilde C_t$ recorded at every gradient update. To reduce fluctuations from finite-batch sampling, we divide the trace into nonoverlapping blocks of $1,000$ updates and compute the median cost $\overline C_k$ within each block. All runs are evaluated at the common reference $T_{\mathrm{ref}}= 500,000$ updates. We define $C_{\mathrm{ref}}$ as the median of $\overline C_k$ over the final $20\%$ of blocks and estimate the residual variability as $1.4826 \times$ the median absolute deviation over the same blocks, denoted by $\widehat\sigma$. The resolved improvement is $\Delta=\overline C_0-C_{\mathrm{ref}}$. If $\Delta\leq3\widehat\sigma$, the improvement is not distinguishable from residual training fluctuations, and we set $t_{\mathrm{conv}}=0$. Otherwise, $t_{\mathrm{conv}}$ is the starting update of the earliest block after which $\overline C_k\leq C_{\mathrm{ref}}+\max(0.05\Delta,\widehat\sigma)$ holds for all subsequent blocks up to $T_{\mathrm{ref}}$. Runs with no qualifying block before the final reference window are classified as non-converged.

\subsection{Compactly encoded spinless Hubbard models and DMRG}
\label{app:hubbard-dmrg}

This subsection provides the numerical details for the Hubbard experiments described in Sec.~\ref{sec:hubbard-results}. For each lattice size, we construct the compactly encoded qubit Hamiltonian;
the reference ground state is obtained separately by applying DMRG to the corresponding Jordan--Wigner representation of the same fermionic Hamiltonian, yielding a matrix-product-state approximation on the $L^2$-qubit register. Expectation values and covariances of operators in the compact encoding are then evaluated by pulling the relevant encoded Pauli operators back to the Jordan--Wigner register and contracting them with this MPS. These MPS-derived moments are used both to evaluate the measurement protocols and to compute the group variances required by the oracle Neyman baseline.

The oracle baseline first partitions the Hamiltonian into a fixed
collection of QWC groups and then allocates the common measurement
budget according to their DMRG-computed variances. The resulting
integer shot counts are used to simulate the corresponding energy
estimator. \model is trained independently using only the encoded
Hamiltonian coefficients and state-independent Pauli-coverage
statistics.

\subsubsection{Compactly encoded spinless Hubbard models}

We consider the spinless Fermi--Hubbard model on $L\times L$ square lattices with open boundary conditions,
\begin{equation}
H_{\mathrm f}
=
-t\sum_{\langle pr\rangle}
\left(
    a_p^\dagger a_r+a_r^\dagger a_p
\right)
+
U\sum_{\langle pr\rangle}n_pn_r
-
\mu\sum_p n_p,
\qquad
n_p=a_p^\dagger a_p,
\label{eq:app-hubbard-fermionic}
\end{equation}
where $\langle pr\rangle$ denotes nearest-neighbour lattice sites. We set $t= 2, U =1, \mu=0.25$ in dimensionless units. The fermionic Hamiltonian is mapped to qubits using the compact encoding ~\cite{derby2021compact,dyrenkova2025scalable}. The construction is conveniently expressed using the two Majorana operators associated with each fermionic mode,
\begin{equation}
    \gamma_p
    =
    a_p+a_p^\dagger,
    \qquad
    \overline{\gamma}_p
    =
    \frac{a_p-a_p^\dagger}{i},
\end{equation}
and the corresponding vertex operators are defined as $V_p=-i\gamma_p\overline{\gamma}_p= I-2n_p$; oriented edge operators are defined as  $E_{pr} = -i\gamma_p\gamma_r$ with $E_{rp}=-E_{pr}$. The encoding assigns low-weight Pauli operators to $V_p$ and $E_{pr}$ while preserving their local commutation and anticommutation relations. Nonlocal fermionic exchange statistics are instead represented by the entangled stabilizer-code subspace.

For the qubit layout, the square-lattice faces are two-coloured in a checkerboard pattern. One vertex qubit is assigned to each fermionic mode, and one auxiliary face qubit is assigned to every face of one colour. Although
the physical Hamiltonian contains only open-boundary bonds, we retain the uniform periodic checkerboard code geometry used in the even-$L$ construction of Ref.~\cite{dyrenkova2025scalable}; no wrap-around hopping or interaction term is included in Eq.~\eqref{eq:app-hubbard-fermionic}. The encoded register therefore contains
\begin{equation}
    n =\frac{3L^2}{2}
\label{eq:app-dk-qubit-count}
\end{equation}
qubits. Thus, the $4\times4$ system uses $24$ qubits, comprising $16$ vertex and $8$ face qubits, whereas the $6\times6$ system uses $54$ qubits, comprising $36$ vertex and $18$ face qubits. The original planar construction requires fewer than $1.5$ qubits per fermionic mode; the periodic checkerboard layout used here has exactly $1.5$ qubits per mode.

The mapped vertex operator is
\begin{equation}
    \widetilde V_p=Z_p.
\end{equation}
An orientation is assigned to every lattice edge such that the edges circulate around alternating checkerboard
faces. For an oriented bond from $p$ to $r$, the mapped edge operator can be written as
\begin{equation}
    \widetilde E_{pr}
    =
    \eta_{pr}X_pY_rF_{pr},
    \qquad
    \widetilde E_{rp}
    =
    -\widetilde E_{pr},
\label{eq:app-dk-edge-map}
\end{equation}
where $\eta_{pr}\in\{-1,+1\}$ contains the orientation sign. The operator $F_{pr}$ acts on the face qubit associated with the bond and is a Pauli-$Y$ operator for a horizontal bond and a Pauli-$X$ operator for a vertical bond, under the convention of Ref.~\cite{dyrenkova2025scalable}.

The local terms of the Hubbard Hamiltonian consequently map to
\begin{align}
n_p
&\mapsto
\frac{1}{2}
\left(
    I-Z_p
\right),
\label{eq:app-dk-number-map}
\\
n_pn_r
&\mapsto
\frac{1}{4}
\left(
    I-Z_p-Z_r+Z_pZ_r
\right),
\label{eq:app-dk-interaction-map}
\\
-t
\left(
    a_p^\dagger a_r+a_r^\dagger a_p
\right)
&\mapsto
-\frac{t}{2}\,
\eta_{pr}
\left(
    X_pX_r+Y_pY_r
\right)
F_{pr}.
\label{eq:app-dk-hopping-map}
\end{align}
The occupation and density--density terms therefore have Pauli weight at most two, while each hopping contribution is the sum of two Pauli strings of weight three. This should be contrasted with the Jordan--Wigner mapping ~\cite{jordan1928paulische} of a two-dimensional lattice, for which geometrically local hopping operators generally acquire strings whose weight grows with the linear system size.

An open $L\times L$ square lattice contains $N_{\mathrm b} = 2L(L-1)$ nearest-neighbor bonds. After combining identical Pauli strings and excluding the identity, the encoded physical Hamiltonian contains one single-$Z$ term per site, one $Z_pZ_r$ term per bond, and two hopping Pauli strings per bond. The number of nonidentity physical Pauli terms is therefore 
\begin{equation}
    M
    =
    L^2+3N_{\mathrm b}
    =
    7L^2-6L.
\label{eq:app-dk-term-count}
\end{equation}
This gives the number of Pauli terms 88, and 216 for $4 \times 4$ and $6 \times 6$ respectively. These are the terms included in the measurement-design objective.

The mapped edge operators must additionally satisfy the fermionic loop identities. These identities are enforced by restricting the qubit state to the common $+1$ eigenspace of a local stabilizer group ~\cite{derby2021compact}. For the checkerboard geometry, we use one loop generator for each face that does not carry a face qubit. If the four surrounding vertex qubits are labelled $p_1,p_2,p_3,p_4$ clockwise, a generator has the form
\begin{equation}
\begin{split}
S_i
={}&
Z_{p_1}Z_{p_2}Z_{p_3}Z_{p_4}
\\
&\times
Y_{f(p_1,p_2)}
Y_{f(p_3,p_4)}
X_{f(p_2,p_3)}
X_{f(p_4,p_1)},
\end{split}
\label{eq:app-dk-stabilizer}
\end{equation}
up to the equivalent indexing convention determined by the edge orientation. Here $f(p_i,p_j)$ denotes the auxiliary face qubit associated with the edge connecting vertices $p_i$ and $p_j$ under the chosen checkerboard encoding.
Each generator has Pauli weight eight, acting on four vertex qubits and four neighboring face qubits~\cite{dyrenkova2025scalable}. We use $8$ such generators for the $4\times4$ lattice and $18$ for the $6\times6$ lattice.

For the ground-state calculation, we energetically select the desired stabilizer sector using 
\begin{equation}
H_{\mathrm{enc}}^{(\lambda)}
=
H_{\mathrm{enc}}
+
\lambda
\sum_{i=1}^{N_s}
\frac{I-S_i}{2},
\qquad
\lambda=6.
\label{eq:app-stabilizer-penalty}
\end{equation}
For each generator, $(I-S_i)/2$ is the projector onto its $-1$ eigenspace. The penalty vanishes identically in the target code space and assigns a positive energy to a violated loop
constraint. 

The stabilizer-penalty terms are excluded from the physical energy estimator, the hit counts, and the state-independent measurement proxy. 
Including the identity and stabilizer penalties, the stored $4\times4$ and $6\times6$ Pauli Hamiltonians contain $97$ and $235$ terms, respectively.

\subsubsection{DMRG and MPS}
\label{app:hubbard-mps}

Although the measurement protocols are defined on the $24$- and $54$-qubit compactly encoded registers, the reference ground states are computed on the corresponding $16$- and $36$-qubit Jordan--Wigner registers.
Let $H_{\mathrm{JW}}^{(L)}$ denote the Jordan--Wigner representation of the spinless Hubbard Hamiltonian in Eq.~\eqref{eq:app-hubbard-fermionic}, with $m=L^2$ qubits. We therefore use $m=16$ for the $4\times4$ lattice and $m=36$ for the $6\times6$ lattice. The fermion-parity sector is chosen to match the sector represented by the compact encoding. 

For each lattice, the ground state is approximated by a matrix-product state
\begin{equation}
\ket{\phi_L}
=
\sum_{z_1,\ldots,z_m}
A^{[1]}_{z_1}
A^{[2]}_{z_2}
\cdots
A^{[m]}_{z_m}
\ket{z_1z_2\cdots z_m},
\label{eq:app-hubbard-mps}
\end{equation}
where $z_p\in\{0,1\}$ and the auxiliary dimensions of the tensors $A^{[p]}_{z_p}$ are bounded by the chosen maximum bond dimension $\chi_L$. The Hamiltonian $H_{\mathrm{JW}}^{(L)}$ is represented exactly as a matrix-product operator constructed from its Pauli decomposition.

We optimize Eq.~\eqref{eq:app-hubbard-mps} using two-site DMRG~\cite{white1992density}. The local effective-Hamiltonian problems are solved using Lanczos iterations, and all tensor network calculations are performed in double complex precision. The bond dimension is increased during the initial sweeps until the maximum value is reached. For the $4\times4$ calculation, we use a maximum bond dimension $\chi_{4\times4}=256$ and at most $40$ complete sweeps. For the $6\times6$ calculation, we use $\chi_{6\times6}=756$ and at most $40$ complete sweeps.

The converged normalized matrix product state (MPS) is retained as the reference state, and its ground-state energy approximation is the variational expectation value
\begin{equation}
    E_{\mathrm{MPS}}^{(L)}
    =
    \bra{\phi_L}
    H_{\mathrm{JW}}^{(L)}
    \ket{\phi_L}.
\label{eq:app-dmrg-energy}
\end{equation}

To evaluate the measurement protocols on the compactly encoded register, we use the pullback induced by the encoding isometry. Let $\mathcal V_L: \left(\mathbb C^2\right)^{\otimes m} \rightarrow \mathcal C_L \subset \left(\mathbb C^2\right)^{\otimes n}$ map the $m$-qubit Jordan--Wigner register to the code space $\mathcal C_L$ of the $n$-qubit compact encoding. The
encoded reference state is formally
\begin{equation}
    \ket{\Psi_L}
    =
    \mathcal V_L\ket{\phi_L},
\label{eq:app-encoded-mps-state}
\end{equation}
but this $24$ or $54$ qubit state is never constructed
explicitly. The compact and Jordan--Wigner representations are related by
\begin{equation}
\mathcal V_L^\dagger
H_{\mathrm{enc}}^{(L)}
\mathcal V_L
=
H_{\mathrm{JW}}^{(L)}.
\label{eq:app-encoded-jw-equivalence}
\end{equation}

Each physical encoded Pauli term $P_k$ preserves the code
space and pulls back to a signed Pauli operator on the
Jordan--Wigner register,
\begin{equation}
    \mathcal V_L^\dagger
    P_k
    \mathcal V_L
    =
    \sigma_k Q_k,
\label{eq:app-hubbard-pullback}
\end{equation}
where $\sigma_k\in\{-1,+1\}$, $Q_k$ is an $m$-qubit Pauli string. For products of
physical terms required by the energy estimator,
\begin{equation}
    \mathcal V_L^\dagger
    P_kP_\ell
    \mathcal V_L
    =
    \sigma_k\sigma_\ell
    Q_kQ_\ell.
\label{eq:app-hubbard-pullback-product}
\end{equation}
The signs and Pauli strings are obtained by reducing the
binary symplectic representation of the encoded operator
over the stabilizer and logical generators of the compact
encoding, while tracking the associated Pauli phase.

The first and second moments entering the measurement
statistics are therefore
\begin{align}
    \mu_k
    &=
    \bra{\Psi_L}P_k\ket{\Psi_L}
    =
    \sigma_k
    \bra{\phi_L}Q_k\ket{\phi_L},
    \label{eq:app-hubbard-mps-first-moment}
    \\
    \Gamma_{k\ell}
    &=
    \bra{\Psi_L}P_kP_\ell\ket{\Psi_L}
    -
    \mu_k\mu_\ell
    \nonumber\\
    &=
    \sigma_k\sigma_\ell
    \bra{\phi_L}Q_kQ_\ell\ket{\phi_L}
    -
    \mu_k\mu_\ell.
    \label{eq:app-hubbard-mps-covariance}
\end{align}
The expectation values on the right-hand sides are evaluated
by direct tensor-network contraction with the $16$- or
$36$-qubit MPS.

This procedure can be viewed as a Heisenberg-picture
evaluation of the measurement circuits. stabilizer tableau
propagation determines which encoded Pauli terms and term
pairs are recovered from each measurement setting. These
operators are then pulled back to the smaller
Jordan--Wigner register before contraction with the MPS.
Consequently, we do not evolve a $24$ or $54$ qubit
encoded MPS separately through every measurement circuit.
The same MPS derived moments are used to evaluate \model and
to compute the group variances of the oracle QWC baseline.

\subsubsection{Hubbard energy root-mean-squared error}
For the Hubbard experiments, the reported error is the conditional root-mean-squared error of a fixed measurement schedule, evaluated directly from the MPS-derived first and second moments. For shot $j$, define its contribution to the FlowMeas energy estimator as

\begin{equation}
    Y_j
    =
    \sum_{k:h_k(\mathbf U)>0}
    \frac{c_k\chi_{jk}}{h_k(\mathbf U)}
    \xi_{jk}.
    \label{eq:app-flowmeas-shot-contribution}
\end{equation}
The complete estimator can then be written as
\begin{equation}
    \widehat E(\mathbf U)
    =
    c_0+\sum_{j=1}^{N}Y_j.
\end{equation}
Because outcomes from different shots are sampled independently, its
conditional variance is
\begin{equation}
\begin{aligned}
    \operatorname{Var}(\mathbf U)
    &=
    \operatorname{Var}_{\rho_L}
    \left[
        \widehat E(\mathbf U)
        \,\middle|\,
        \mathbf U
    \right]
    \\
    &=
    \sum_{j=1}^{N}
    \sum_{\substack{k,\ell:
    h_k(\mathbf U)>0,\,
    h_\ell(\mathbf U)>0}}
    \frac{
        c_kc_\ell
        \chi_{jk}\chi_{j\ell}
    }{
        h_k(\mathbf U)h_\ell(\mathbf U)
    }
    \Gamma_{k\ell},
\end{aligned}
\label{eq:app-flowmeas-hubbard-variance}
\end{equation}
where only Pauli pairs simultaneously covered by circuit $U_j$ contribute to the inner sum.

If some physical Pauli terms remain uncovered, the resulting bias is
\begin{equation}
    \operatorname{Bias}(\mathbf U)
    =
    -
    \sum_{k:h_k(\mathbf U)=0}
    c_k\mu_k.
    \label{eq:app-flowmeas-hubbard-bias}
\end{equation}
The reported FlowMeas root-mean-squared error is therefore
\begin{equation}
    \operatorname{RMSE}(\mathbf U)
    =
    \sqrt{
        \operatorname{Var}(\mathbf U)
        +
        \operatorname{Bias}(\mathbf U)^2
    }.
    \label{eq:app-flowmeas-hubbard-rmse}
\end{equation}
All sums in these expressions contain only the physical Pauli terms of
the compactly encoded Hubbard Hamiltonian. Identity and
stabilizer-penalty terms are excluded.

\subsubsection{QWC baseline for spinless Hubbard model}
\label{app:hubbard_qwc}
The Hubbard comparison uses a common measurement budget of $N = 1000$. For \model, the reported schedules use a maximum CNOT depth $d_{\max}=2$ and are trained with the state-independent DSS proxy in Eq.~\ref{eq:cost_dss}. \model training uses only the encoded Hamiltonian coefficients and state-independent Pauli-coverage statistics through the terminal reward.

The oracle baseline first partitions the physical encoded Hamiltonian into QWC groups using sorted insertion
~\cite{crawford2021efficient}. The Pauli terms are ordered by decreasing $|c_k|$ and inserted sequentially into the first existing group whose accumulated product basis is compatible with the new term on every qubit. A new group is opened when no compatible group exists. This procedure produces QWC groups. 

For a QWC group $g$, the variance of its single-shot contribution to the energy is
\begin{equation}
    V_g
    =
    \sum_{k,\ell\in g}
    c_kc_\ell\Gamma_{k\ell}.
\label{eq:app-qwc-group-variance}
\end{equation}

If the group is measured $N_g$ times, the conditional variance of the resulting energy estimator is
\begin{equation}
    \operatorname{Var}_{\mathrm{QWC}}
    =
    \sum_g\frac{V_g}{N_g},
\label{eq:app-qwc-variance}
\end{equation}
where $\sum_gN_g=N$. Minimizing this expression under the fixed total budget gives the continuous Neyman allocation
\begin{equation}
    N_g^\star
    =
    N
    \frac{\sqrt{V_g}}
    {\displaystyle\sum_{g'}\sqrt{V_{g'}}}.
\label{eq:app-neyman-allocation}
\end{equation}
The corresponding minimum variance is
\begin{equation}
    \operatorname{Var}_{\mathrm{QWC}}^\star
    =
    \frac{1}{N}
    \left(
        \sum_g\sqrt{V_g}
    \right)^2.
\label{eq:app-neyman-variance}
\end{equation}
This is the globally variance-minimizing continuous allocation for the fixed grouping. Integer shot counts are obtained by largest-remainder rounding, with at least one shot assigned to every group and with the final counts constrained to sum exactly to $N$. Since $V_g$ is computed from the target-state covariances, this is an oracle allocation and uses information that is unavailable to the state-independent \model protocol.

Because the QWC partition covers every physical Pauli term, its estimator is unbiased. Its reported root-mean-squared error is
\begin{equation}
    \operatorname{RMSE}_{\mathrm{QWC}}
    =
    \sqrt{
        \sum_g\frac{V_g}{N_g}
    },
    \label{eq:app-qwc-hubbard-rmse}
\end{equation}
where $N_g$ denotes the final integer shot allocation.

\subsection{Energy-estimation protocol}
\label{app:energy-evaluation}

For a fixed measurement schedule, each numerical trial independently samples the corresponding measurement outcomes from the same reference state. For the molecular experiments, the reported root-mean-squared error is
\begin{equation}
\operatorname{RMSE}
=
\sqrt{
\frac{1}{500}
\sum_{r=1}^{500}
\left(\widehat E^{(r)}-E_0\right)^2
},
\label{eq:app-rmse}
\end{equation}
where $E_0$ is the reference ground-state energy including the identity shift. The 500 trials quantify shot noise for a fixed schedule; they do not quantify variability across independent training runs. If a Pauli term remains uncovered, its contribution is assigned the zero estimator as described in Sec.~\ref{sec:methods}.

For the direct DSS comparison in Fig.~\ref{fig:dss}, we use the mean absolute error reported by DSS under a budget of 1000 measurements and one CNOT layer. The DSS values are taken from Ref.~\cite{kirk_derandomized_2024}; they are not recomputed. The \model comparison uses the same DSS cost in Eq.~\eqref{eq:cost_dss}.
\subsection{Computational throughput benchmark}
\label{app:throughput}

The throughput benchmark includes neural-network inference, masked action sampling, Clifford propagation, hit-matrix construction, proxy and reward evaluation, trajectory-balance loss evaluation, backpropagation, and the Adam update. Compilation warm-up is discarded. The reported time per update is averaged over 10,000 updates, from iteration 1000 through iteration 11000, on one NVIDIA H100 GPU.

\begin{figure}[htbp]
\centering
\includegraphics[width=0.48\linewidth]{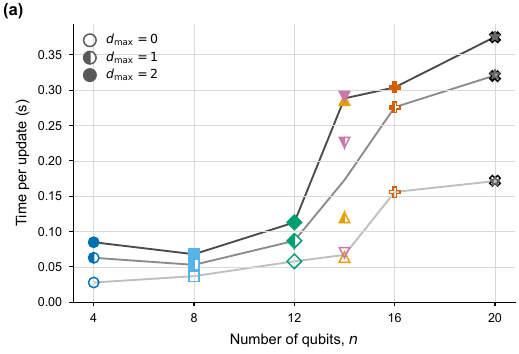}
\includegraphics[width=0.48\linewidth]{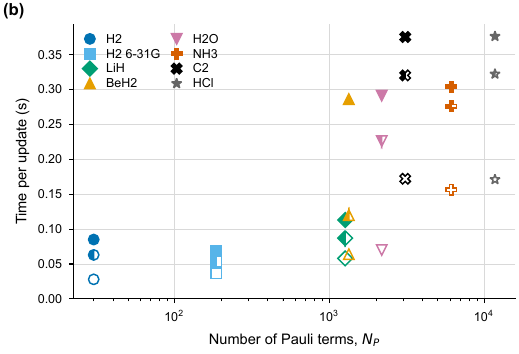}
\caption{\textbf{Training throughput for the molecular benchmarks.} Time per gradient update as a function of \textbf{(a)} the number of qubits and \textbf{(b)} the number of Pauli terms. Marker fill indicates the CNOT-depth budget in panel \textbf{(a)}, and marker shape identifies the molecular Hamiltonian in panel \textbf{(b)}. Each update includes the complete training pipeline described in the text.}
\label{fig:training-throughput}
\end{figure}

The per-update benchmark does not by itself determine the total classical design cost, which also depends on the number of updates required for convergence. For the molecular benchmarks, the measured per-update times translate into modest end-to-end costs. For a full training campaign of $5\times10^{5}$ updates, the cost ranges from approximately four GPU-hours for H$_2$ at depth budget 0 ($0.028$\,s per update) to $2.2$ GPU-days for HCl at depth budget 2 ($0.376$\,s per update) on a single H100. The scaling is markedly sublinear in Hamiltonian size. Across a $390$-fold increase in the number of Pauli terms, the per-update time grows by only a factor of $4$--$6$, so even the largest molecular Hamiltonians considered here remain well within single-GPU reach.

The same accounting applies to the two-dimensional Hubbard experiments, whose
wall-clock cost we measured directly from production training runs on a single NVIDIA H100 GPU, averaging the interval between logged updates over the full course of each run. We note that the $6{\times}6$ experiments use a batch size of 8 elements per update due to GPU
memory constraints (Table~\ref{tab:training-hyperparameters}). One gradient update takes $0.53$, $0.63$, and $0.74$\,s on the $4{\times}4$ lattice (24 qubits) at CNOT-depth budgets of 0, 1, and 2, respectively, and $0.51$, $0.86$, and $1.15$\,s on the $6{\times}6$ lattice (54 qubits). A full campaign of $10^{6}$ updates therefore requires $6.1$--$8.5$ GPU-days for the 24-qubit system and $5.9$--$13.3$ GPU-days for the 54-qubit system, depending on the depth budget. Notably, at depth budget 0 the 54-qubit system is no more expensive per update than the 24-qubit system ($\approx 0.5$\,s), as fixed per-update costs dominate; the gap opens only at larger depth budgets, where the deeper circuits on the larger lattice yield proportionally longer trajectories.

\end{document}